**Manuscript for**

# Short and Long-range cyclic patterns in flows of DNA solutions in microfluidic obstacle arrays

Oskar E. Ström, Jason P. Beech, Jonas O. Tegenfeldt(*)

Division of Solid State Physics, Department of Physics, Lund University

NanoLund, Lund University.

*Jonas O. Tegenfeldt.

**Email:** jonas.tegenfeldt@ftf.lth.se

**Abstract**

We observe regular patterns emerging across multiple length scales with high-concentration DNA solutions in microfluidic pillar arrays at low Reynolds numbers and high Deborah. Interacting vortices between pillars lead to long-range order in the form of large travelling waves consisting of DNA at high concentration and extension. Waves are formed in quadratic arrays of pillars, while randomizing the position of the pillar in each unit cell of a quadratic array leads to suppression of the long-range patterns. We find that concentrations exceeding the overlap concentration of the DNA enables the waves, and exploring the behavior of the waves as a function of flow rate, buffer composition, concentration and molecular length, we identify elastic effects as central to the origin of the waves. Our work may not only help increase the low throughput that often limits sample processing in microfluidics, it may also provide a platform for further studies of the underlying viscoelastic mechanisms.

## Main Text

### Introduction

While the ability to analyze small samples using small reagent volumes is often considered a main advantage of microfluidics, low analyte throughput can also limit the types of studies that can be performed. If useful throughput is the amount of analyte per unit time, then it depends on both the volume throughput of the carrying fluid, and the concentration of the analyte. It can be increased by increasing either or both of these factors. However, different microfluidic phenomena occur at different ranges of flow velocities and concentrations, which places method-dependent constraints on the useful parts of the parameter space. As an example, we can consider Deterministic Lateral Displacement (DLD)[1].

Recently, high throughput DNA sorting was demonstrated using DLD [2]. Throughputs reached 24 µL/h and 760 ng/h which corresponds to orders of magnitude improvement over existing microfluidic approaches. DLD has also been used for sorting of physiological concentrations of blood [3] at high throughputs [4]. Due to the nature of the sample both for blood and for DNA solutions, further increase in throughput is expected to be hampered by viscoelastic effects. While viscoelastic fluids have been shown to have a tunable effect in DLD [5], we showed both positive and negative contributions to the separation of long (>20 kbp) DNA molecules at high analyte concentrations. An observation we made during these separations at high DNA concentrations or high flow velocities was the emergence of fluctuations in concentration and in the local flow direction that led to decreases in the quality of the separation at concentrations of tens to hundreds of µg/mL[2]. Another example where high DNA concentrations are relevant is for sample preparation for nanopore sequencing (e.g. Oxford Nanopore: Ligation Sequencing Kit (XL SQK-LSK109-XL and Rapid Sequencing Kit (SQK-RAD004)). The standard protocols require up to approximately 50 μg/mL which is within the range of concentrations for which we observe waves.

Here, we study the conditions under which these fluctuations emerge as striking patterns of macroscopic ordered waves in the rapid flow of high concentrations of DNA across our devices, which consist of quadratic arrays of micrometer sized pillars. Although fluctuations are common in these systems, refer to the recent review paper[6] by Datta *et al.* for a thorough introduction to viscoelastic flow phenomena, our waves are strikingly regular and long range. The waves consist of DNA at higher concentration and extension, with two specific orientations and flow directions mirrored along the central axis of the device. We study their behavior as a function of flow rate, buffer composition, concentration and molecular length by direct observation of the DNA using both fluorescence and polarization techniques. In the flow of the DNA through our devices, we observe regions where the DNA appears to move in circular trajectories. We refer to these regions as vortices. We find that several phenomena occur at different spatial and temporal scales. At the single pillar scale ($10^{-5}$ m), vortices form, grow, interact, and are shed into the flow. At the array scale ($10^{-3}$ m), fluctuations in concentration are seen to form and to travel through the array. Based on the observed phenomena at multiple scales, we propose some possible mechanisms for the formation of the waves. We note that the formation of bands of DNA, depletion zones and vortices precede the emergence of waves. The synchronization of fluctuations and the shedding of vortices coincides with the passing wave fronts.

To describe the fundamental underlying mechanisms that lead to the formation of the macroscopic waves, we turn to various dimensionless numbers. Here, it is important to keep in mind that the dimensionless numbers and concentration ratios presented in this work are only useful for representing the state of the system at the onset of instability and wave behavior. Specifically, during states characterized by strong fluctuations, these numbers vary considerably with time and with location in the device.

We calculate the Reynolds number and find that $Re = \rho \cdot u \cdot w / \eta_s < 0.1$ , where $\rho$ and $\eta_s$ is the density and viscosity of the solvent, $u$ is the mean flow velocity and $w$ is the gap between array rows, thereby excluding inertial effects in our systems. The Deborah number, $De = (u/L_{pp})\tau_Z$, where $L_{pp}$ is center-to-center distance between array rows and $\tau_Z$ is the Zimm relaxation time, and the Elasticity number, $El = De/Re$, are then used to quantify elastic effects. In control experiments we obtained a relaxation time for the highest concentration that was consistent with the estimate based on the Zimm relaxation time, see Supporting Information, section 4. As a simplification and as a conservative estimate, we define *De*, *Re*, and thus *El* as independent of concentration but remind the reader that polymer relaxation time has been shown to vary strongly with concentration [7-11], the type of flow [11], the shear rate [12] and displays a high degree of molecular individualism, especially in the entangled regime [8,9]. To illustrate the role of interactions between DNA molecules, we refer to *C/C** which is the ratio between the concentration, *C*, of the DNA solution and the overlap concentration, $C^* = M/[(4\pi/3)R_g^3 N_A]$ [13]. Here, *M* is the molecular weight of the DNA and $N_A$ is Avogadro's number. *C** is the concentration above which DNA molecules no longer behave like isolated, individual molecules. We modify *C** by changing $R_g$ through changes in DNA molecular length and ionic strength. Intermolecular interactions range in magnitude from the semi-dilute unentangled regime at $C^*$ to the "entangled" at $C_e \approx 3C^*$ (lowest reported limit [9]).

For detailed calculations and discussions about the dimensionless numbers, see Supporting Information, section 3.

In the following, we will refer to columns and rows of pillars. Rows are perpendicular to, and columns are parallel with, the flow direction.

**Results**

We note that the flow behavior differs at different length scales. In the following we therefore present our observations of fluorescently stained DNA flowing through a quadratic array first at the macroscopic scale where the waves are visible. We subsequently gradually increase the magnification to investigate the behavior of the flow around a few pillars and finally the flow around single pillars. We use both standard epifluorescence microscopy to visualize the motion of the DNA and fluorescence polarization microscopy to visualize the orientation and the extension of the DNA qualitatively.

We investigate the criteria for wave formation, specifically the role of viscoelastic effects by varying the Deborah number, by varying the flow rates, and the role of intermolecular interactions by varying the ratio between the concentration and the overlap concentration. This is done by simply changing the concentration of the DNA or by manipulating the overlap concentration through varying the composition of the DNA solution.

*Macroscopic view – waves*

Above a threshold flow velocity, we observe long range waves as shown in Figure 1(b), Supporting Movie S1. The transition from uniform flow to waves can be clearly seen when ramping the pressure up and down so that waves emerge and then disperse [Figure 5(a), Supporting Movie S2]. Waves arise simultaneously across the array when high $\Delta p$ is applied and disperse when $\Delta p$ is decreased. The emergence of waves does not occur immediately at the start if the pillar array, but rather at about 10 pillar rows or 180 μm into the array [see Figure 1(b)]. The waves are moving from top to bottom with the fluid flow direction, see Supporting Movie S1. The wave fronts alternate between two orientations (32.8° ± 4.2° and -34.5° ± 7.5° relative to the long-axis of the array) and persist for between hundreds of micrometers to several millimeters of travel along the device. When a large wave of a particular orientation passes or disperses, a large wave of the alternative orientation typically arises in its wake. The DNA waves represent long-range temporal and spatial synchronization of DNA concentrations and orientations across the pillar array. When observing entire wave systems at low magnification (2–4×), we typically observe the intensity to vary (relative to the mean intensity at low flow velocity) from as much as approximately double the mean intensity in the brightest peaks to as low as approximately one third in the darkest troughs. Note that the leading edge of the wave is steeper than the trailing edge.

The occurrence of waves is strongly dependent on the regular geometry of the array. With a small randomization of pillar positions in the quadratic array, no waves are observed at any flow velocities, see Figure 1(c) and Figure 8 and Supporting Movie S2.

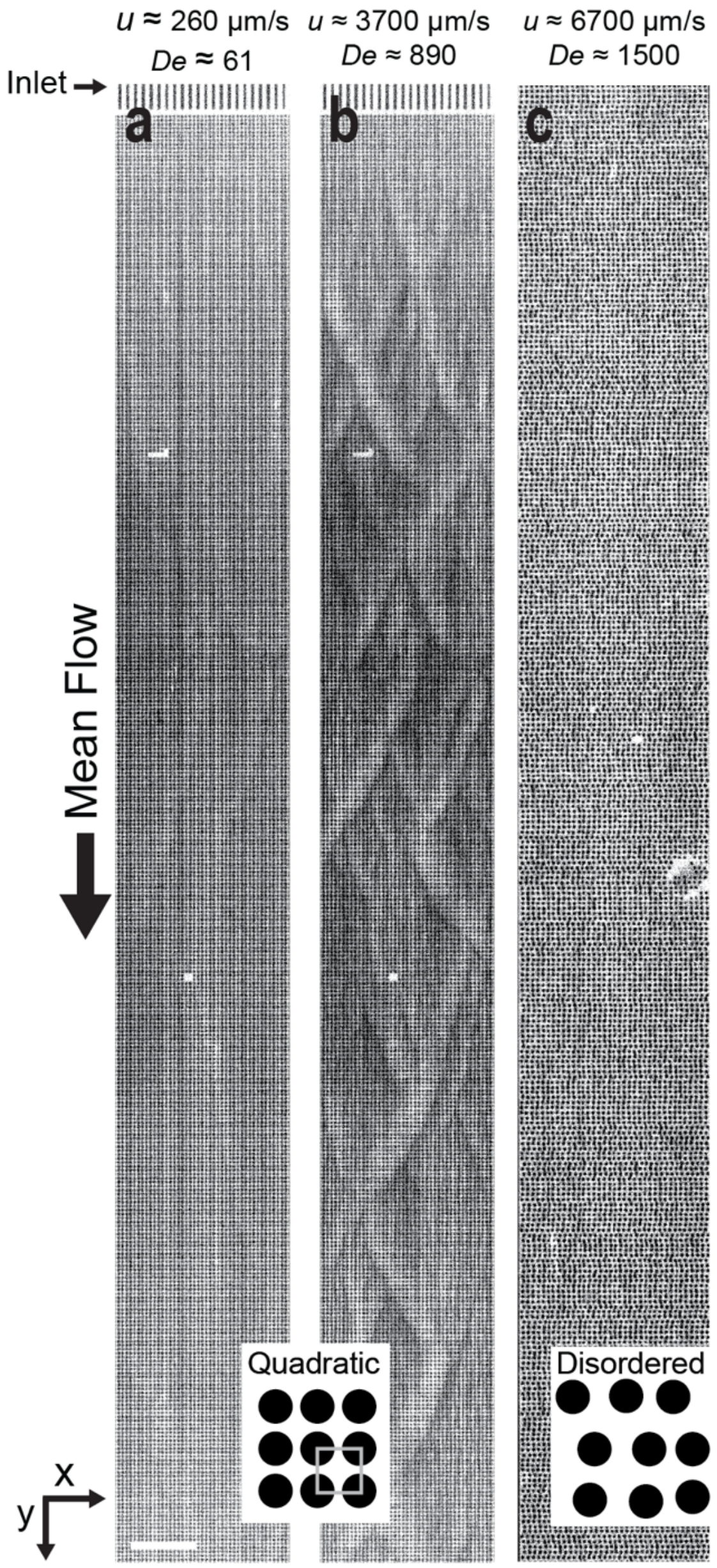

**Figure 1.** Waves emerge as DNA solutions flow above a threshold velocity in a quadratic micropillar-array (b) but not below the threshold (a) or at any flow velocity in a disordered array (c). (a–c) show micrographs acquired with low magnification (2×) of λ DNA solutions at 400 µg/mL). The pressure drop across the array is from top to bottom in the images. See Supporting Movie S1 for a video with the raw data. Scale bar is 300 µm.

*Microscopic view – vortices, depletion zones and flow direction switching*

At greater magnification, see Figure 2, we observe the DNA being depleted between the pillars within the rows as well as pairs of vortices being formed. The flow is seen to change direction, in this case moving from the right to the left as the vortex pair is shed into the flow. This process can be followed in a kymograph, Figure 2(b), showing the fluorescence signal along the red dashed line in Figure 2(a) over time. Note that the vortices become visible only when they contain DNA.

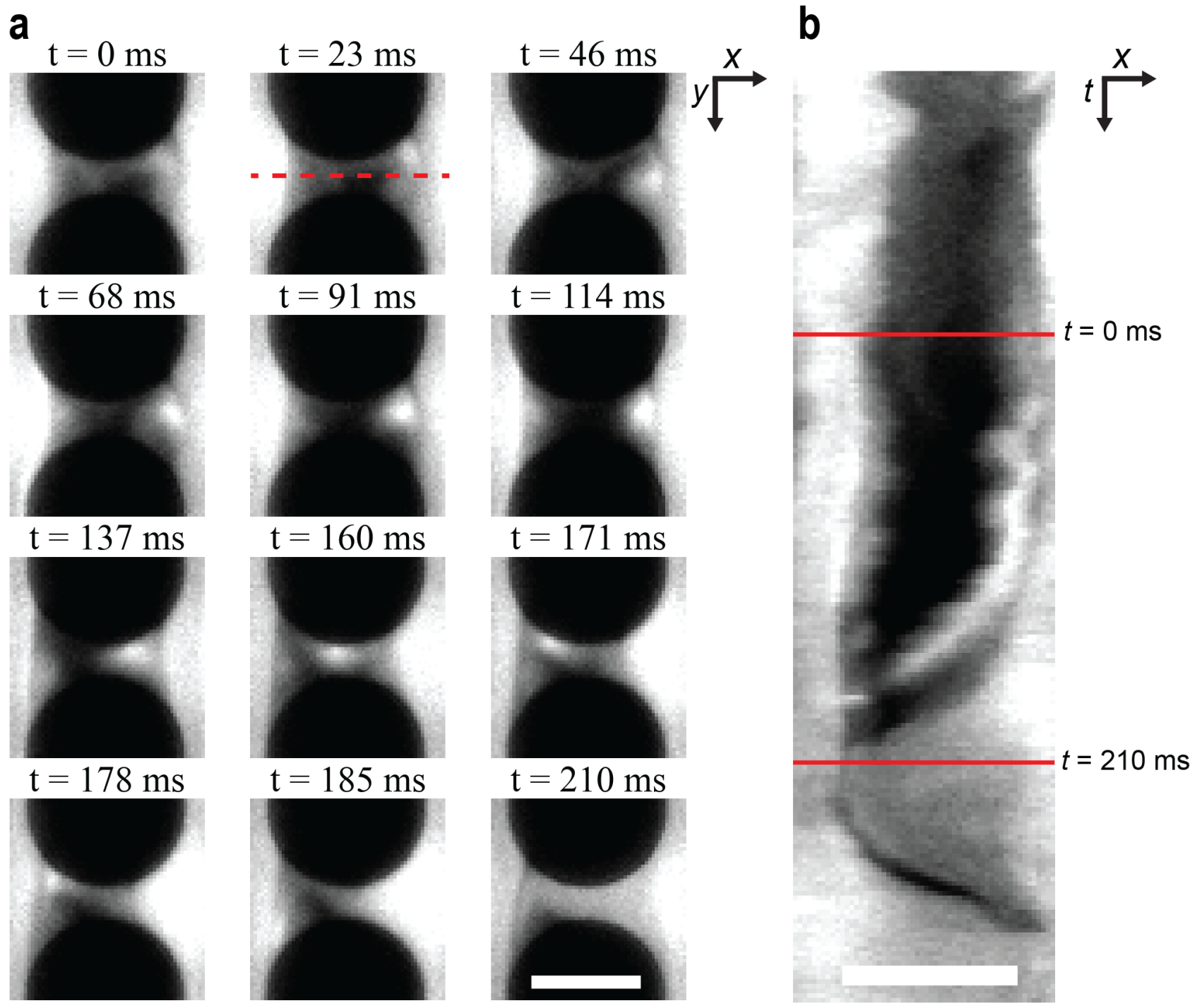

**Figure 2.** Vortex growth and shedding. (a) Time series of a vortex collecting DNA at the right side of the space in between the two pillars. Thereafter, it shifts position to the left side and is subsequently shed. (b) Kymograph over the series of events in a based on the red dashed line. λ DNA solution was used at C = 400 µg/mL, $u \approx 3.2$ mm/s and $De \approx 1300$. See Supporting Movie S3 for a video representing the raw data. Scale bars are 10 µm.

*Intermediate scale – Coupling macroscopic and microscopic behavior*

At intermediate magnification it is possible to observe both the behavior of the DNA around single pillars and the coupling of the dynamic behavior across multiple pillars. Figure 3 shows three regimes that occur, where different phenomena dominate, as the flow velocity is increased. We define the different regimes as follows: Steady depletion (S) - a steady non-fluctuating zone is formed between the pillars. Cyclic (C) - the DNA concentration in the zone between the pillars varies in a cyclic manner with time. Waves (W) - while the microscopic picture appears chaotic, the term alludes to the fact that macroscopically, this is where we see the waves.

At low applied pressure ($u \approx 120$ µm/s, $De \approx 51$), Figure 3(a and d), the DNA becomes depleted between the pillars along the columns. This results in bands of DNA that flow between the columns. At higher applied pressure ($u \approx 650$ µm/s, $De \approx 260$), Figure 3(b and e), the widths of these bands fluctuate and vortices are observed. The kymograph in Figre 3(e) clearly shows both the periodic fluctuation of the widths of the bands of DNA and also how neighboring bands are coupled. As in Figure 2, vortices can be seen. However, under these conditions they persist for longer and can be seen to move back and forth within the depleted zones with the same frequency as the band fluctuations before eventually shedding. At higher applied pressure ($u \approx 3.2$ mm/s, $De \approx 1300$), Figure 3(c and f), waves can be seen. At the wave fronts the flow is seen to switch direction, from moving solely along columns to moving with an additional considerable component along the rows. The orientation of the wave front is perpendicular to this flow direction. This is further corroborated by the polarization measurements that show extension of DNA molecules perpendicular to the wave fronts, see Figure 4. Waves traveling in two directions with orientations approximately mirrored around the overall flow direction, see Figure 1(b), form across the entirety of the device with the waves being coherent for most of their transit through the array.

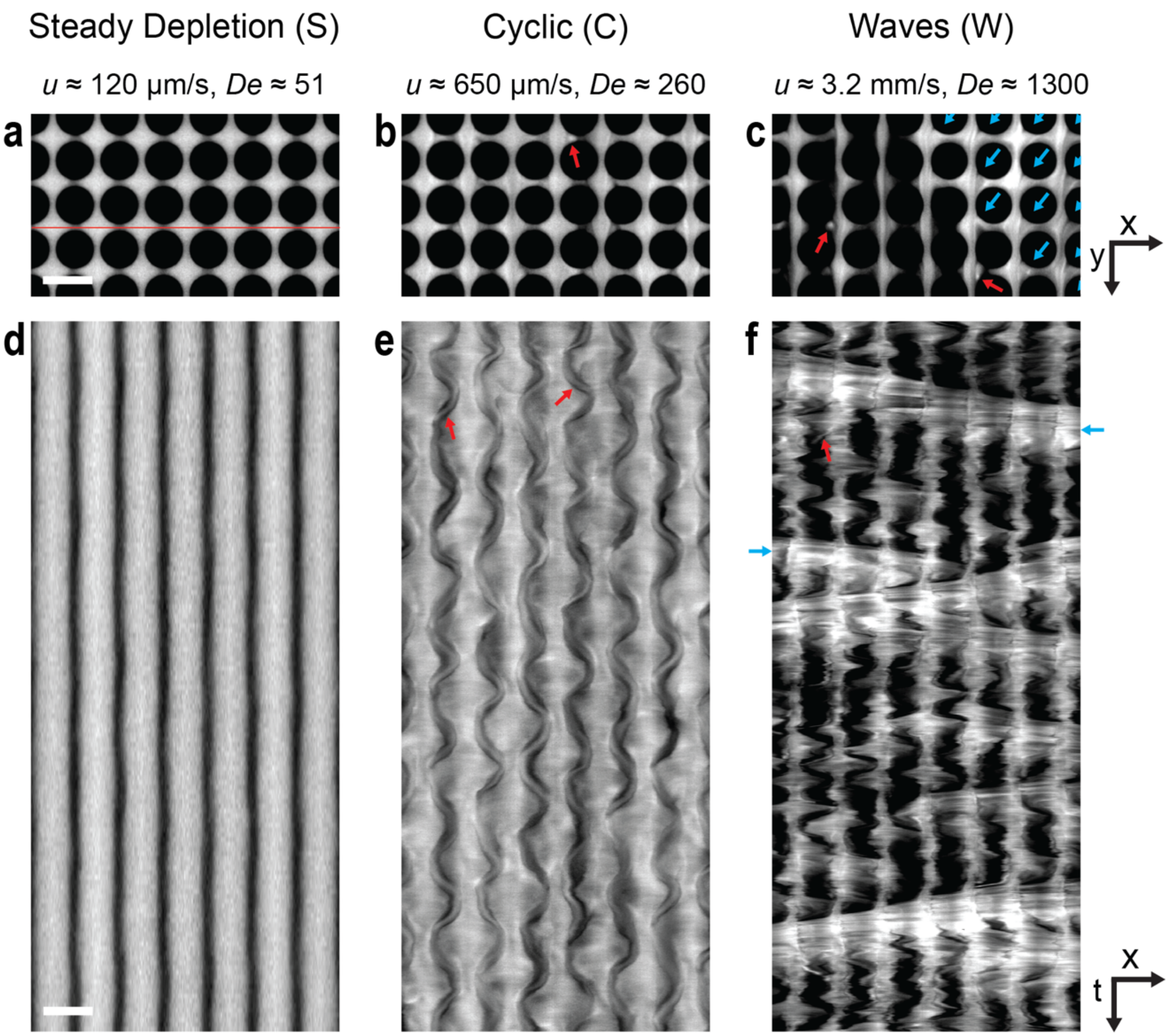

**Figure 3.** Microscale observations of the three regimes: Steady depletion (S) cyclic (C) and waves (W). (a-c) show micrograph snapshots of λ DNA solution at $C$ = 400 μg/mL and 7 × 4 pillars whereas (d-f) shows 4 s long kymographs between the rows in a (see red line). The first regime (S, at low $u$) is characterized by prominent depleted zones in between the pillars along the flow direction with minimal lateral flow fluctuations. The second regime (C) consists of large lateral undulating fluctuations and periodic growth and shedding of blobs in the depleted zones (red arrows). These blobs typically shift from one side of the depleted zone to the other as shown in (b, C regimes). The third regime (W) is defined by oblique waves on-top of large lateral flow fluctuations in-between the waves. In a wave (blue arrows), the depleted zones have shifted to an angle perpendicular to the wave front as seen in a, W). The same contrast and brightness settings have been applied for all images. To enhance visibility at high frame rates, an increased staining ratio of 1 fluorophore to 10 base pairs has been utilized. See Supporting Movie S3 for a video representing the raw data. Scale bars are 300 μm.

*Orientation of DNA Strands*

Polarization measurements of the fluorescence allow us to monitor the local orientation of the DNA micro- as well as macroscopically, because the intercalating dye molecules (YOYO-1) adopt an orientation such that the transition dipole moment of the dye is approximately orthogonal to the backbone of the DNA [14,15]. We excite using unpolarized light and by detecting the polarization of the emission we gain information about the local orientation and stretching of the backbone. Specifically, we image the DNA at two different perpendicular polarization orientations simultaneously, chosen to maximize their difference in intensity. To visualize the polarization on-top of the fluorescence intensity, images are colorized, with the pixel value denoting the fluorescence intensity and the hue denoting the emission polarization ratio, $P = (I_{\parallel} - I_{\perp})/(I_{\parallel} + I_{\perp})$, see Figure 4. See the Materials and Methods for details on the image processing. At high magnification (100×) we observe two important phenomena. Firstly, the wave propagation orientation varies between the two wave angles and is rarely parallel to the side walls of the channel as would be expected for motion of fluids for Newtonian fluids or for viscoelastic fluids in the S regime. Secondly, the molecules are concentrated in streaks around the pillars [Figure 4(d and f), Supporting Movie S6] and are stretched in the flow direction. At lower magnification (10×) we see that the polarization of DNA is the same throughout a wave and that the DNA is stretched perpendicularly to the orientation of the wave fronts, see Figure 4(a and b) and Supporting Movie S7.

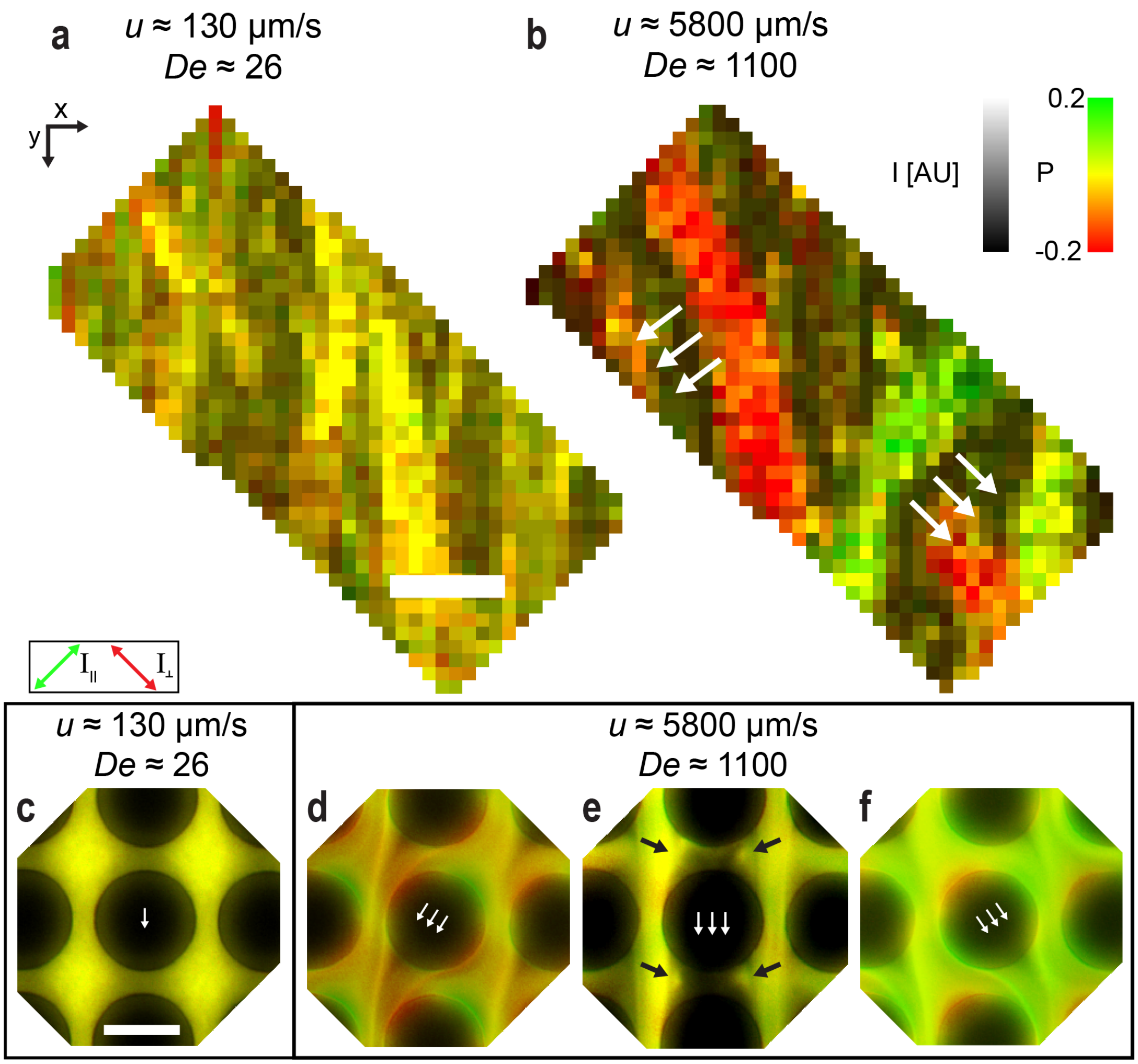

**Figure 4.** Polarization imaging showing the absence of polymer extension for low flow velocities and the significant extension along the wave fronts for high flow velocities. The signal appears green when the DNA molecules orient with one polarization channel and red with the other, perpendicular channel. (a and b) Snapshots of pillar arrays at low magnification (10×) and imaged processed to remove the pillars. Sample is 400 µg/mL λ DNA. The images are colored where the fluorescence intensity, *I*, is represented by the pixel value and polarization emission ratio, *P*, is represented by the hue, saturation and value (HSV) color model. (c-f) Snapshots at high magnification (100×) for the same experimental run as (a and b). DNA strands are concentrated into diagonal streaks and there is a significant number of extended polymers that are pressed into the leading edges of the pillars, resulting in a contrasting polarization ratio. The black arrows in panel e point to the vortices. See Supporting Movies S6 and S7 for videos representing the raw data. Scale bar is 200 µm in (a) and 10 µm in (c).

*Effect of Flow Velocity*

To illustrate the influence of elastic effects we vary the velocity, thereby varying *De*, by changing the applied pressure. The progression S-C-W-C-S is clearly shown in Figure 5(a) where $\Delta p$ is ramped from zero to a value above which waves are seen and back.

The kymographs in Figure 4(c) show the waves with different time scales such that the total distance shown is similar for each flow velocity. It is clear from here that the lengths scales of the spatial pattern are the same independent of the flow velocity.

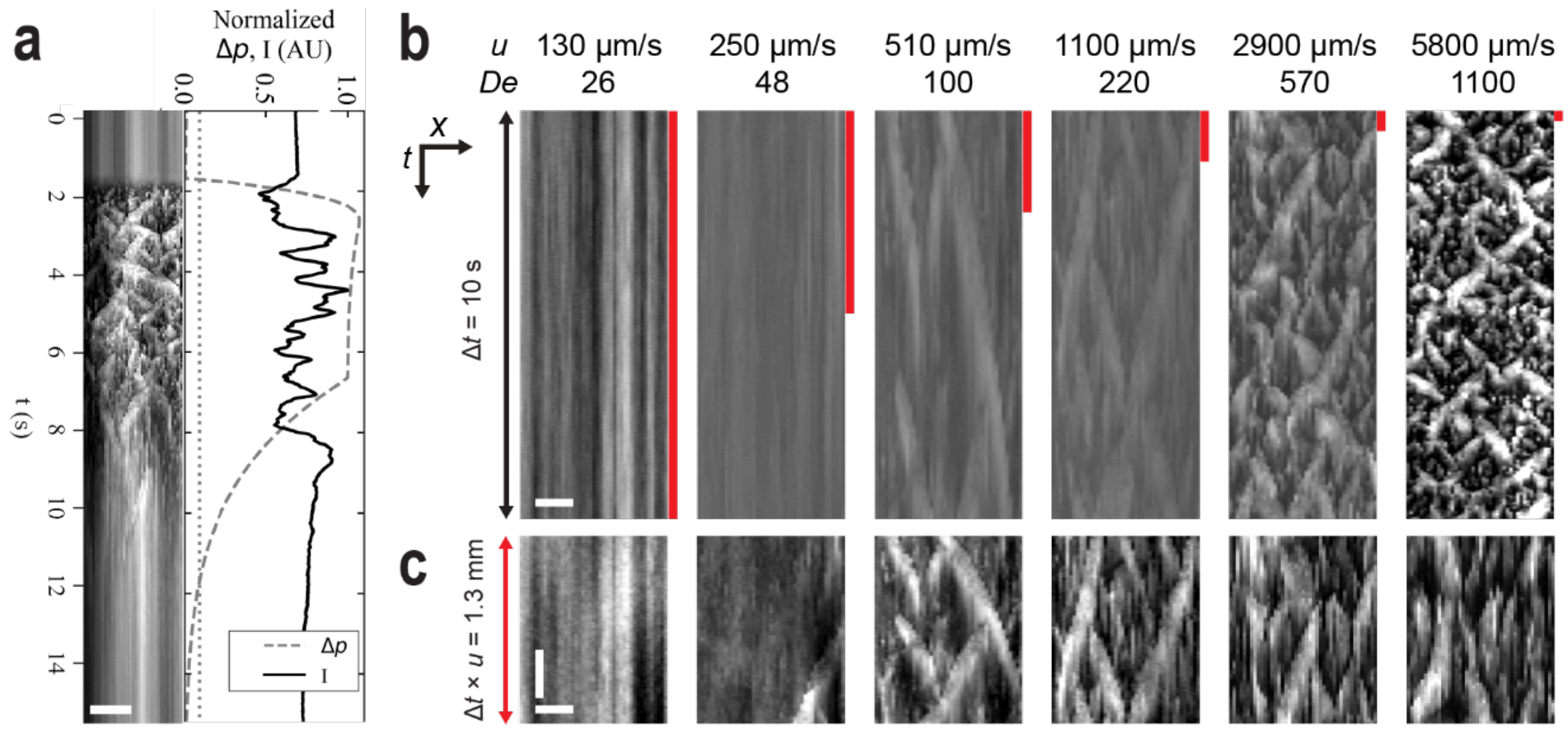

**Figure 5.** Wave formation as a function of flow velocity for a λ DNA solution at 400 μg/mL. The pressure drop across the array is from top to bottom in the images. (a) Kymograph showing how waves appear on the application of a high pressure difference, $\Delta p$ = 1 bar ($u \approx$ 5.8 mm/s and $De \approx$ 1,100 at maximum $\Delta p$), with the wave pattern dissipating when $\Delta p$ is removed. The kymograph is based on image data acquired at 10× magnification of a pillar row half-way across an array, where each pixel value corresponds to the mean fluorescence intensity in the upstream regions of each pillar. Alongside the kymograph, $\Delta p$ as a dashed line, and the mean intensity of 5 × 5 upstream regions for each time point, solid line, are plotted. The dotted line corresponds to the non-fluorescent background intensity. (b) Kymographs of a cross-section of the array show clear waves at higher *u* and *De*. Movies have been captured with a frame rate of 47.5 fps so that each of the kymographs in (b) represent 10 s. The brightness and contrast settings are the same for each kymograph. The data from (c) scaled such that the presented distance is the same for all flow velocities. This is the result of scaling the time span, $\Delta t$, for each of the kymographs to the respective flow velocity. The time duration for each flow velocity in (c) is related to the kymographs in (b) with red vertical lines. Brightness and contrast settings have been set independently for each kymograph to highlight the similarity in the wave pattern. The horizontal scale bars are 300 μm in (a) and 200 μm in (b and c), and the vertical scale bar in (c) is 2 mm.

*Effect of DNA Concentration*

To elucidate the effect of inter-molecular interactions we varied the concentration of DNA. We repeat experiments with λ DNA at concentrations between 25 μg/mL << *C** and 422 μg/mL >> *C**. The lower limit was set at a concentration for which no significant instabilities were observed, and the upper limit was set by the concentration of the available DNA stock solution. See Table 1 for details on the preparation of the DNA samples. For solutions of λ DNA (48.5 kbp) with *C*/*C** << 1, a transition from S to C is observed for flow velocities of approximately 600 μm/s which corresponds to *De* ≈ 130, see Figure 6(a). When C/C* ≈ 1, C appears earlier, 100 μm/s and *De* ≈ 25, and a further transition to W is observed at 7 mm/s and *De* ≈ $1 \times 10^3$. For the highest concentration tested, transitions from S to C and then to W are observed already for *De* < $10^3$ and flow velocity of < 3 mm/s. When *C*/*C** > 1, the relative magnitude of the elastic properties of the DNA solution dominates and relaxation times become much longer than the transit times across the curved streamlines around the pillars. See Supporting Movies S4a and S4b for a video comparison between low and high *C* at high *De*.

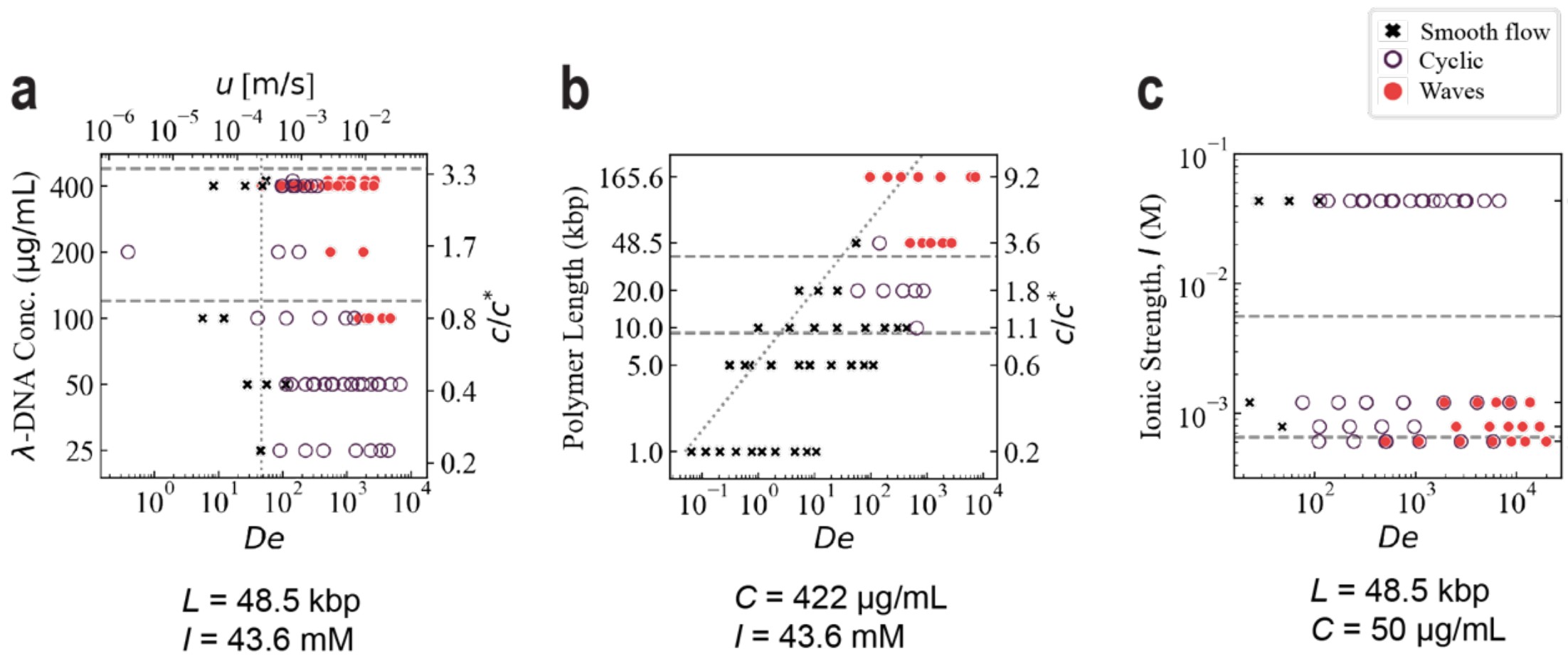

**Figure 6.** Phase diagrams of occurrence of waves in the quadratic array as a function of (a) polymer concentration of λ DNA, (b) polymer length at constant concentration (422 μg/mL) and (c) ionic strength of the sample for λ DNA at 50 μg/mL. The dashed horizontal lines indicate $C^*$ and the lower entanglement limit, $C_e = 3C^*$ [9]. All three diagrams illustrate the requirements of high $C/C^*$ and high flow velocities for waves to arise. $u$ represents the estimated mean flow velocity at the gaps of the array. To visualize the dependency of $De$ on polymer length and ionic strength, respectively, the dotted non-horizontal lines in the three panels denote the lowest velocity where waves have been observed for a λ DNA sample, $u$ = 250 μm/s. The shape of these lines represents the dependency of $De$ on polymer length and ionic strength, respectively. The number of independent data points is 226. In all three panels, the axes are scaled logarithmically except for the $C/C^*$ values indicated in the right-hand side axes which represent the individual values for the corresponding data sets and do not follow a simple scaling relationship. Note that some data points are labelled as exhibiting both a cyclic flow and waves. This represents data acquired in different devices. Note also that the data points denoted as Smooth flow represent both entirely homogenous distribution of DNA as well as the Steady Depletion (S) as depicted in Figure 3(a, d).

### *Effect of Polymer Length*

To further probe the influence of intermolecular interactions, we manipulate the overlap concentration by varying the contour length of the DNA. Note that this also affects the relaxation time. In experiments with DNA of different contour lengths we find that at the same concentration ($C$ = 422 μg/mL) only solutions of long molecules exhibit waves, see Figure 6(a). See Table 2 for listing of the molecular lengths used and their physical parameters. The shortest samples, 1 kbp and 5 kbp, exhibit no flow fluctuations or waves (S regime) even at the highest velocities. Similarly to the results of varying the concentration, we start to observe the appearance of waves at $C/C^* \approx O(1)$.

*Effect of Ionic Strength*

Another way to manipulate the overlap concentration, we vary the salt concentration and thereby the swelling of the polymer due to the varying electrostatic interactions. We find that, when below the threshold concentration of DNA for wave formation, lowering the ionic strength of the sample, *I*, causes waves to form. See Table 1 for a detailed list of the tested samples of varied *I*. DNA is a polyelectrolyte with phosphate groups on the backbone conferring a negative charge. Increasing the ionic strength screens this charge, decreases the inter-polymer interactions and decreases $R_g$, taking the solution further away from the overlap concentration. Our most prominent example of the effect can be seen in the figure for λ DNA at 50 µg/mL [Figure 6(c)]. Waves are clearly seen at low ionic strengths but disappear completely at higher ionic strengths. It is interesting to note that the waves appearing under these conditions are slightly different than those at higher salt concentrations. Waves of different orientations occur only in separate parts of the array and do not cross, see Supporting Movie S5.

*Effect of Solvent Viscosity*

To explore whether shear thinning plays an essential role in the formation of waves, we start with low concentration of DNA and add sucrose to adjust the viscosity to the values used above. The viscosity of the DNA solution is dependent on the DNA concentration and DNA solutions are also known to be highly shear thinning[16,17]. The addition of sucrose increases the viscosity and is also known to reduce the shear thinning of DNA solutions giving us a Boger fluid (elastic fluid with constant viscosity) [18]. Supporting

Figure S1 shows that, for λ DNA at 50 μg/mL in 43% sucrose, clear waves form whereas at similar *De* with the DNA only dissolved in buffer, no waves appear. Even at *De* up to $1 \times 10^4$, no wave formation can be observed without added sucrose. Adding sucrose decreases the threshold concentration of DNA over which we observe waves at a given flow velocity. In other words, exceeding the overlap concentration is not a necessary condition for the initiation of the waves.

**Discussion**

We demonstrate ordered macroscopic waves across device-scale distances which are the result of the synchronization of microscopic flow patterns around microscopic pillars.

*Fluctuations and elastic turbulence in viscoelastic flows*

Several examples of elastic turbulence in micropillar arrays have previously been reported in the literature. One example occurring at high *Wi (corresponding to the Deborah number used in our work)* is the asymmetric transfer of fluid across single [19,20], pairs [21] or an entire array of micro-pillars [22]. For the case of single pillar or pair of pillars, the transfer of fluids is seen to be stable over time for certain ranges of *Wi* while for the array case, the "lane change" occurs transiently in aperiodic ways at random locations. This phenomena is similar to flow velocity waves that have been reported to propagate between the neighboring streams of pillar-pairs [23]. Another phenomenon is the observation of coherent structures or streaks at flow velocities of similar magnitude. These have been shown to self-organize in straight channels in the wake following either 1D [24] or 2D arrays [25] of pillars. In the latter case, the streaks propagate perpendicularly to the flow in cycles, and eventually disintegrate at the end of each cycle. We believe these phenomena to be closely related to the large-scale waves presented here.

Waves with somewhat similar appearance as those in our work have been observed recently[26] in the flow of viscoelastic media in an array of pillars, so called canopy waves, but the geometry of the array, the molecules involved, the method of observation and the conclusions drawn about the nature of the waves differ from those presented here. Instead of fluorescently labelled DNA, the authors use hyaluronic acid combined with beads to trace the flow. Both flexible and rigid pillars are used with a slender geometry and, in contrast to our device, with a gap between the top of the pillar and the device cover so that the flow takes place not only between but also above the pillars. While our waves are characterized primarily by different concentrations and alignment of the DNA, the canopy waves are defined by distinct magnitudes and directions of the velocity vector field of the flow as well as motion of the pillars.

To our knowledge, DNA solutions have only been used as a sample to study elastic instabilities in microfluidics for contraction-flow devices [27-29]. Otherwise, a plethora of various solutions of other molecules have been used. These include solutions of aqueous wormlike micelles (WLM) [19,21,30-33], high-weight polyacrylamide (pAAm) [34,35], polyacrylamide in glycerol [36,37], hydrolyzed polyacrylamide (HPAM) [22,38], hyaluronic acid[26], or polyethylene oxide (PEO) in water [39-41] or in glycerol [42-44].

Almost all of the published work on viscoelastic effects in porous materials has focused on either larger pillar or pillar gap dimensions. Arrays of pillars of the dimensions reported in these previous studies have little relevance for the separation of DNA molecules using microfluidics and specifically DLD. The combination of the small length-scales [$R \approx 7$ µm, $w$ (gap) $\approx 4$ µm] together with the long relaxation times of long DNA strands puts our work in a regime of ultra-low *Re* and high *De* and thus ultra-high

*El*. The smaller length-scale also results in smaller radius of curvature of the streamlines that has long been known to enhance elastic turbulence in a range of flow geometries[41,45,46].

We estimate *De* in our experiments to cover a range between $10^{-1}$ and $10^{4}$ across all studied samples and flow rates. *Re* ranges instead between $10^{-8}$ and $10^{-3}$ (the upper limit based on the viscosity of the solvent) and thus *El* reaches $10^{5}$ - $10^{9}$. Waves appear at *De* approximately $10^{2}$ – $10^{3}$ for concentrated λ DNA and T4 DNA and around *De* approximately $10^{3}$–$10^{4}$ for the low salt (0.1× TE, 0.13× TE and 0.2× TE) and low-concentration (50 µg/mL) λ DNA samples. Interestingly, $El \approx 10^{8}$ for all occurrences of waves. The banding that we observe occurs for even the lowest flow velocities that we were able to observe but not at zero flow. This indicates the importance of elastic forces even at vanishingly low flow velocities in our system.

Our results are in agreement with studies showing an increased degree of elastic effects with reduced salt [38,47] and polymer concentrations [28]. Interestingly, Gulati *et al.* observed no vortex formation at 40 µg/mL λ DNA but increasing vortex formation with higher *Wi (corresponding to the Deborah number used in our work)* for 400 µg/mL λ DNA, similar to the lack of waves for our 50 µg/mL sample but waves occurring at $C \geq 100$ µg/mL. Note that the onset of elastic turbulence has been shown to be independent of the polymer concentration in large micropillar arrays [48] and occurring at lower *Wi* for lower concentration for a polydisperse DNA solution in a micro-contraction geometry [27].

*Emergence of waves*

We believe that waves form when a critical threshold of elastic stresses is exceeded. The threshold depends strongly on *C/C** such that higher *C/C** gives stronger intermolecular interactions which in turn increases the relaxation times. At higher flow velocities, shear rates are higher, leading to greater DNA extension and less time for the DNA molecules to relax as they flow across the curved streamlines. This in turn leads to higher rates of elastic stresses, resulting in temporal and spatial fluctuations in DNA concentration, local flow direction and in turn on the orientation, and extension of molecules. Others have shown similar effects in various microfluidic geometries where the degree of spatio-temporal flow velocity fluctuations has increased with increasing Weissenberg number, $Wi$ [19,22,23,27-31,34-41,43,44,49]. In our array, these instabilities present as bands with fluctuating widths and as vortex pairs that periodically grow and shed blobs of polymers. Similar processes have been observed previously for other polymer solutions [38,44] although the synchronization of these fluctuations into large-scale wave patterns has not.

Based on our imaging of DNA flow at multiple scales and on correlations between the observed phenomena we propose some possible mechanisms that might underlie the formation of waves.

*Microscopic vortices, depletion zones and lift forces*

On the microscopic scale of the individual pillar, depletion zones are formed between the pillars, which leads to bands of DNA between columns [Figure 2 and Figure 3(a and d)]. These might have their origins in interaction forces between the DNA and the pillars. Close to the channel wall, the number of possible conformations for a polymer is lower and entropic effects lead to migration away from the wall on the scale of the radius of gyration. This effect is not dependent on flow and might be considerable at very low flow rates leading to the depletion of DNA molecules from that part of the solvent that flows into the slow-moving regions between the pillars. During flow, the hydrodynamic wall lift force [50] and/or the elastic lift force [51] might push the DNA away from the pillar wall and contribute to depletion. The elastic lift force is predicted to increase with the polymer concentration and flow velocity [52]. Further studies are required to ascertain which, if any, of these effects underlie the formation of depletion zones.

At higher flow rates, and larger *De*, which depends on many factors such as the length of the DNA and the concentration as described above, the time required for the DNA to change its conformation is long compared to the interaction time with the pillars and elastic stresses build up. While we are unable currently to prove causation, we do see correlation between the formation of vortex pairs in the depletion zones and the oscillating widths of the DNA bands that form between the pillar columns, both of which become more pronounced as *De* increases. Once formed, vortices begin to pull in passing DNA and the concentration increases several-fold. When the vortex pair is “full” of DNA, it leaves the depletion zone and flows downstream (Figure 2). Due to viscous coupling and large local changes in the flow field due to the vortex and emerging DNA “blob”, neighboring vortex pairs can be triggered to shed if they are critical in this sense. This could explain some of the concentration, and other threshold behavior of wave formation since it is only large systems of unstable vortex pairs that form simultaneously that can synchronize their shedding into waves. Subcritical systems form vortices but not at sufficient densities for waves to form. We believe that the nature of our periodic vortex pair growth and shedding is of similar nature to what has been previously reported in the literature [38,44]. Note however, for the reported shedding reported in a 2D pillar array [38], there were large spacings between the pillars (porosity, $\phi = V_{pore}/V_{total} \approx 0.7$, compared to $\phi \approx 0.52$ in our array), and thus less blocking of their pillar gaps occur by the shed polymer blob. We speculate that the smaller spacings in our array could lead to a higher degree of coupling between neighboring vortices.

At sufficiently high *De,* the coupled fluctuations in the flow along the columns and the formation and shedding of vortices is unable to dissipate the buildup of elastic stress and large-scale changes in the flow direction are triggered as DNA flows between the columns. This flow direction switching occurs simultaneously with the shedding of DNA filled vortices over large numbers of pillars and is seen as a wave moving along the device. Because flow with a component towards one side of the channel will lead to a buildup of pressure, the local flow direction must switch, and so the next wave is oriented in the opposite direction.

Polarization measurements provide strong evidence for the relationship between the microscopic phenomena that emerge around the pillars and the formation of large-scale waves. At high magnification (100×) we see that vortex shedding takes place in either of two directions. As the DNA strands become stretched, the degree of polarization increases, which we illustrate as a clear red or green signal in Figure 4. The fact that at low magnification (10×) entire waves are seen to contain DNA with the same orientation, and that this is the same orientation that is observed at the microscopic level indicates that it is the process of synchronized, directional vortex shedding that constitutes waves. The individual waves correspond not only to a higher concentration but also to a higher degree of stretching compared to the surrounding bulk DNA.

From Supporting Movie S5, acquired at low magnification (4×), it is clear that some of the waves radiate from specific sites in the array. This is consistent with the waves emerging from a system of inherently unstable vortices. When the conditions for wave emergence are reached, waves may be nucleated by random fluctuations, by the passing of a previous wave, or by a defect or clogging event in the array. Exactly why fixed nucleation points are more prevalent for some conditions is yet to be discerned.

*Effect of solvent viscosity and shear thinning - Boger fluids*

To elucidate the role of shear thinning we add sucrose to raise the solvent viscosity, leading to wave formation at lower C/C*. We note that because of reduced shear thinning due to the added sucrose, the viscosity and thus the relaxation time stays high even at high flow velocity. Elastic fluids with constant viscosity are commonly known as 'Boger fluids' [53] and have been widely employed in the study of elastic turbulence in microfluidic systems [36,37,42-44]. As a consequence of the added viscosity, the elastic stresses, represented by *El*, are increased by two orders of magnitude (from approximately $4.7 \times 10^7$ without sucrose to approximately $2.0 \times 10^9$ with sucrose). Our observations are in agreement with Hemminger *et al.* who also observed increased elastic instabilities when adding sucrose to a DNA solution. They flowed calf thymus DNA solution through a micro-contraction and observed an upstream vortex growth with 40% (w/w) sucrose but not without, at similar *De* [27]. Similarly as when reducing the length scale, a significant increase in the relative magnitude in the elastic forces is achieved when increasing the solvent viscosity, which in turn could explain wave emergence with added sucrose.

*Relationship between length scales of array and polymers*

The lengths of the DNA molecules may contribute in different ways to the formation of waves. Other than molecule-molecule interactions, as quantified using C/C*, the size of the molecules relative to the array geometry is posited to be a decisive factor. We find that it is only the long molecules that have a contour length on the same or greater length scale as the pitch of the array (~18 μm) that can form waves. As these molecules may span between a flow constriction and a flow expansion site and can also become wrapped around pillars (see the work by [54] and [55]), they couple these together via both their interactions with the pillars, the fluid and with one another, amplifying the elastic nature of the fluid. We show this wrapping occurring in our devices in Figure 4(d and f), where the polarization is perpendicular to the flow lines directly upstream of the pillars. This is consistent with the observation by Shi and coworkers of stronger viscoelastic fluctuations at similar *De* for pillars that were more tightly spaced in the flow direction in a 1D pillar array [42]. In our case, the pillars are very tightly spaced which is likely to lead to a strong conformational hysteresis for the polymers when they flow across multiple pores.

## Conclusions & Outlook

We have observed the formation of device-scale waves in polymer solutions moving through a micropillar array. We identify the waves to consist of areas of stretched and oriented polymers, at higher concentrations than the bulk average, moving at two well-defined angles symmetrically around the mean flow direction. We have shown the link between the depletion of DNA between pillars, the generation and shedding of elastic vortex pairs, and the emergence of large-scale waves when fluctuations in these processes become synchronized. From the set of parameters commonly varied in microfluidic work with polymer solutions we have identified crucial factors involved in the formation of waves in micropillar arrays.

The prospects to perform microfluidic operations on polymer solutions such as DNA that are relevant for a wide range of applications, at much higher concentrations than those presented in the literature, relies on an understanding of the onset of potentially disruptive phenomena such as those presented here. This may be particularly relevant for handling of complex biological samples consisting of high concentrations of macromolecules of different compositions exhibiting viscoelastic effects. From the perspective of separations in deterministic lateral displacement devices, this knowledge may allow throughputs to reach orders of magnitude larger per device cross-sectional area (1–200 μg/h in the present work) than previously shown. Conversely, mixing, which is another important microfluidic unit operation, is increased at the onset of flow instabilities [56]. Understanding the phenomena presented here may facilitate the optimized mixing of analytes at high concentration.

In the interest of better understanding the fundamental mechanisms involved, in the future it might be of interest to carefully characterize the extension of the DNA in order to, for example, quantify the elastic energy flow in and out of the polymer contributing to the viscoelastic effects. From a more practical point of view, it would also be interesting to further explore the effects of array geometry as well as temporal and spatial fluctuations *e.g.,* using multiple coloring schemes, particle image velocimetry (PIV) and frequency analysis. We showed how a randomization of the pillar array suppresses the waves. However, randomization can take place in many different ways that would be worth exploring. In our work we have not studied in detail the nucleation of the waves.

An improved understanding of these types of mechanisms would be highly relevant for future device designs where waves are leveraged as a mechanism for improved mixing or the suppression of waves is used to ensure optimized sorting.

**Materials and Methods**

Our experiments take place in standard microfluidics devices made using soft lithography. A schematic of a device is shown in Figure 9. The pillar arrays are approximately 8 mm long and 800 μm wide and 10.9 ± 0.1 μm in depth. In the quadratic and disordered arrays, the pillars are 14 ± 0.3 μm in diameter. In the quadratic array, the gap is 4 ± 0.3 μm between rows and columns of the array whereas in the disordered array, the gap varies between 3 μm and 6 μm. The data is acquired using standard epi-fluorescence microscopy. The DNA strands are labelled with the bisintercalating fluorescent dye YOYO-1 using standard protocols. For the polarization microscopy, an Optosplit II is used to enable simultaneous acquisition of the two perpendicular polarization channels. Detailed descriptions are found in the Supporting Information: calculations of overlap concentrations and the ionic strengths, discussions and calculations of polymer radii, and calculations and reference measurements of relaxation times.

*Device Design*

The microfluidic devices were designed with the layout editing software L-Edit 16.02 (Tanner Research, Monrovia, CA, USA). The unit cell of the quadratic array is shown in Figure 7.

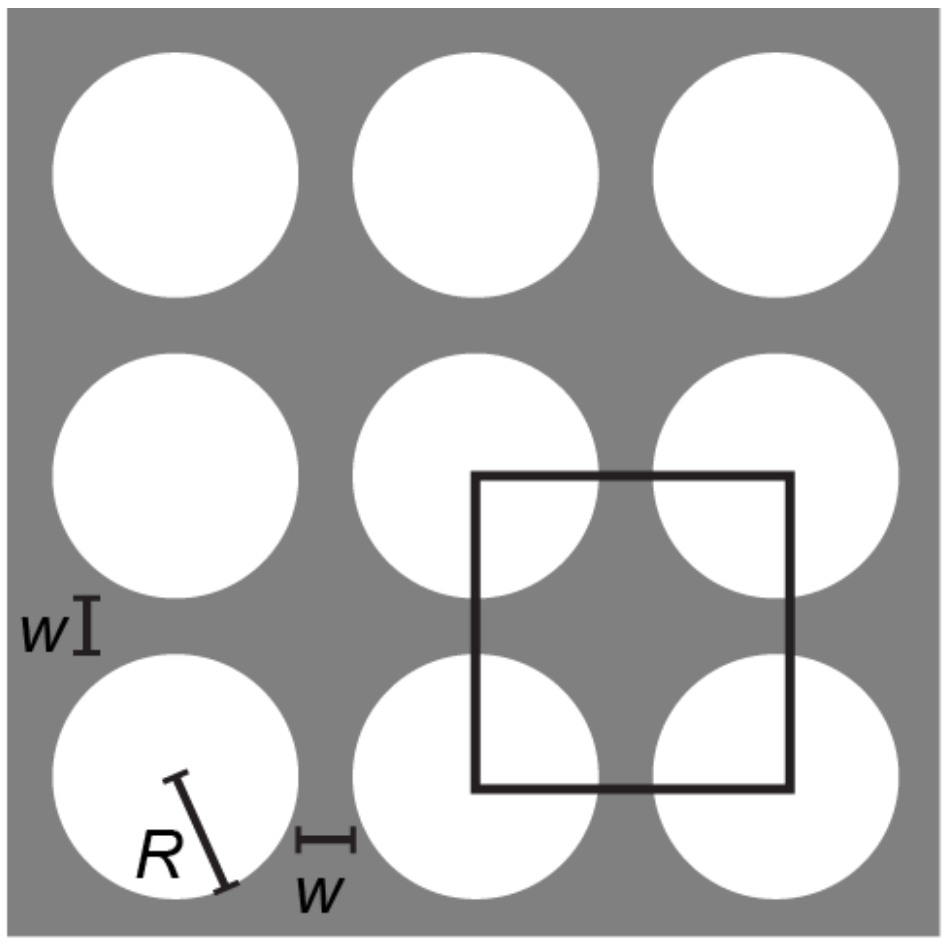

**Figure 7.** Array unit cell designs of the quadratic array. The lateral and vertical pillar-pillar gap, $w$ = 4 $\pm$ 0.3 µm, and the pillar radius, $R$ = 7 $\pm$ 0.15 µm where $\pm$ denotes the standard deviation.

The disordered array was designed to have a random gap size ranging from 3 µm to 6 µm with a random row shift, see Figure 8. The pillars have a radius of $R$ = 7 $\pm$ 0.1 µm. Similarly to the quadratic array, the row-row pitch is set to 11 µm. Every row shifts laterally in a random magnitude between 0 µm and $w_R$ = 18.5 µm. Additionally, the position of the pillar inside each unit cell is randomly placed. The angle from the center of the unit cell, $\theta$, as well as the displacement, $d$, is varied between -180° to 180° and 0 to 1.5 µm, respectively. This arrangement leads to a minimum pillar-pillar gap of 3 µm. This is close to the spatial resolution of the mask aligner with a slight margin.

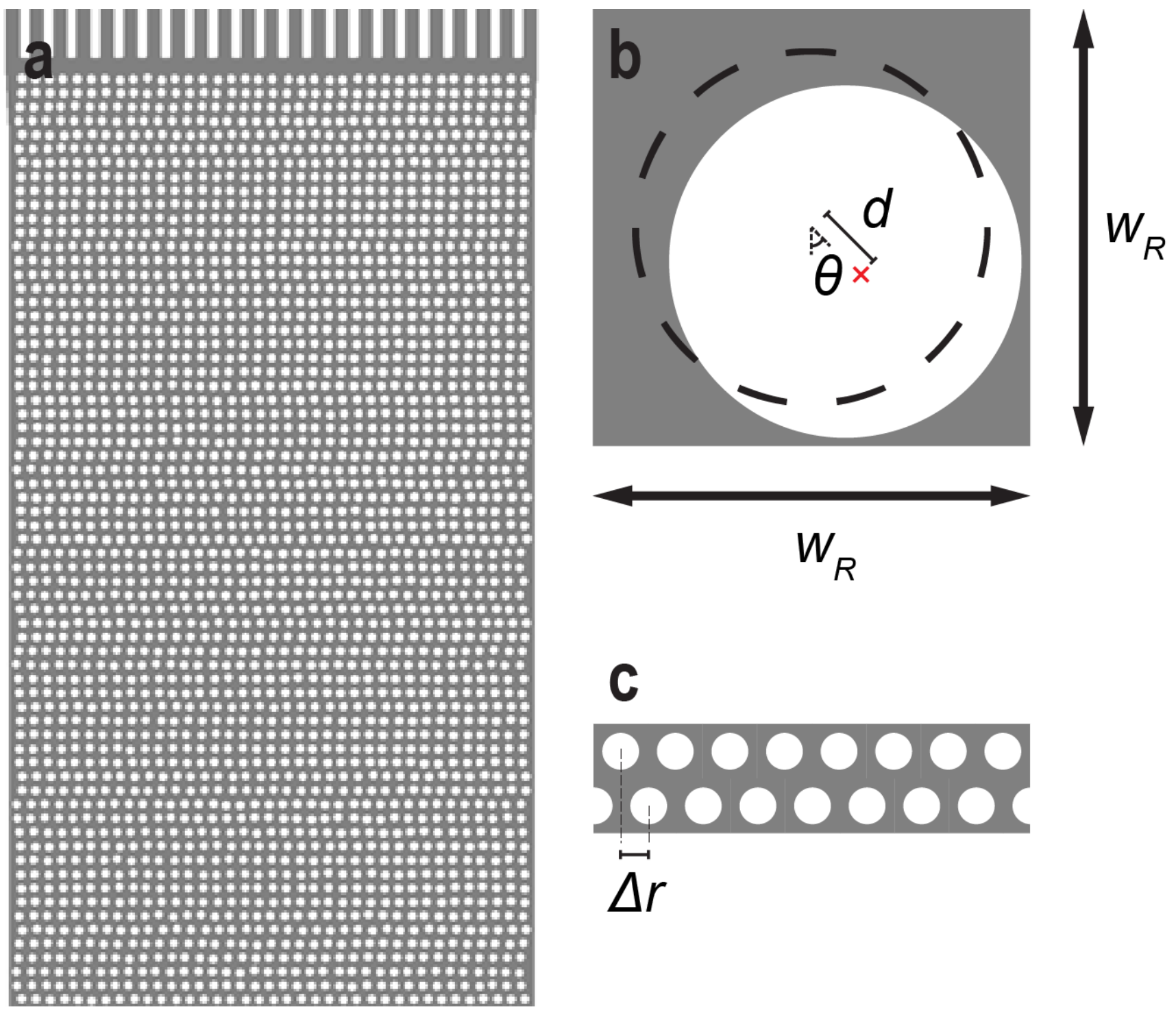

**Figure 8.** Disordered Array Design. (a) shows the design of the inlet region. (b) displays a schematic of a unit cell with width, $w_R$, with displacement, $d$, of the pillar from the unit cell center with a displacement angle, $\theta$. (c) shows a schematic of two rows whereas the lower row has been shifted a distance $\Delta r$.

*Device Fabrication*

The device design was made using the layout editing software L-Edit 16.02 (Tanner Research, Monrovia, CA, USA). The devices were fabricated in polydimethylsiloxane (PDMS, Sylgard 184, Dow Corning, Midland, MI, USA) using standard replica molding [57]. The master mold was made in SU-8 2015 (MicroChem, Newton, MA, USA) using UV-lithography (Karl Süss MJB4, Munich, Germany) with a photomask from Delta Mask (Delta Mask, Enschede, The Netherlands). Directly following the UV-lithography, the mold was coated with an anti-sticking layer of 1H,1H,2H,2H-perfluorooctyltrichlorosilane (ABCR GmbH & Co. KG, Karlsruhe, Germany). PDMS was poured on top of the master mold to a thickness of approximately 6 mm and cured for 1 h at 80 °C. The devices were cut out with a scalpel, holes, 1.5 mm in diameter, were punched for fluidics connection. The resulting PDMS cast was thoroughly rinsed with isopropanol, ultra-purified water (Milli-Q® water, Merck KGaA, Darmstadt, Germany) and blow-dried with nitrogen gas. To evaporate any liquid residue, PDMS casts were then placed on a hotplate (150 °C) for at least 5 min. The same washing procedure was performed with the coverslip glass slides (50 × 24 mm, No. 1.5H, Marienfeld, Lauda-Königshofen, Germany) used to seal the devices. Air plasma (Zepto, Diener electronic GmbH & Co. KG, Ebhausen, Germany) was used to activate PDMS and glass surfaces. Firstly, the glass coverslips were subject to plasma for 90 s. Then, the PDMS slabs were added to the plasma oven and both surfaces were subject to an additional 10 s of plasma. Prior to plasma treatment, the PDMS surfaces were cleaned using adhesive tape (Tesafilm® invisible tape, Tesa AB, Kungsbacka, Sweden). Immediately after plasma exposure, the two surfaces were bonded by pressing the PDMS on top of the glass slide. The device was then put in an oven at 120 °C for roughly a day to render the surface hydrophobic and stored in ambient conditions according to [58] until use. A typical device mounted on the microscope can be seen in Figure 9(b).

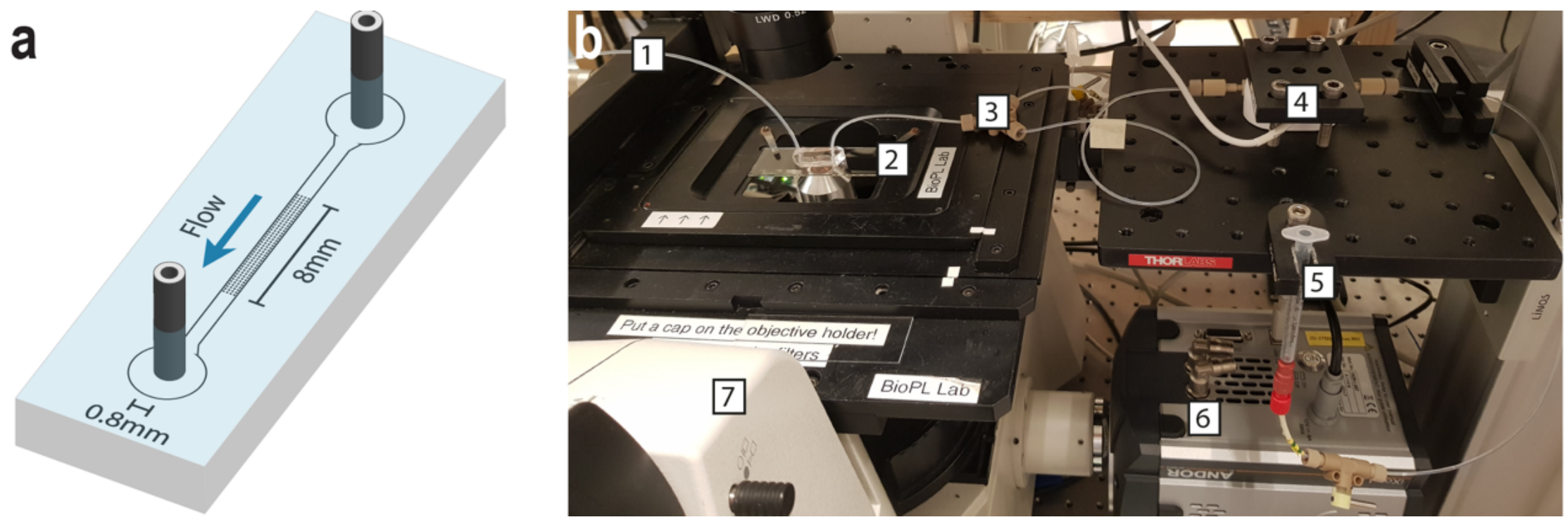

**Figure 9.** Schematics of the device (a) and photograph of the typical setup with the microfluidic device mounted on the epi-fluorescent microscope (b). Pressure difference is applied using a pressure controller with a tubing into the inlet (**1**). The PDMS device (**2**) is mounted on the epifluorescence microscope (**7**). The device is connected with a tubing to a three-way connection that allows for sample discharge into a waste reservoir (**3**) as needed. During operation a valve is used to shut off the waste reservoir. The fluid is then pushed through the flow sensor (**4**) and into the final reservoir (**5**). (**6**) shows an EMCCD-camera used for capturing most of the fluorescence videographs.

### *Experimental Setup*

Flow in devices was generated using nitrogen overpressure which was either controlled with a MFCS-4C pressure controller (Fluigent, Paris, France) or a custom-built manifold when applying pressures over 1 bar. The pressure in the custom-built setup was measured using a manometer (model 840081, Sper Scientific, Scottsdale, AZ, USA). The flow was measured by connecting the outlet to a flow sensor (Flow rate platform with flow unit S, Fluigent, Paris, France). See Figure 10 for graphs of the flow rate dependence on the pressure difference. The devices were imaged using an Eclipse Ti microscope (Nikon Corporation, Tokyo, Japan) with acquisition software NIS-elements AR (v. 5.02.03) with a FITC filter cube, and illuminated with Solis-470C High-Power LED (Thorlabs, Newton, NJ, USA) or SOLA Light Engine (6-LCR-SB, Lumencor Inc, Beaverton, OR, USA). Objectives 2× (Nikon Plan UW, NA 0.06), 4× (Nikon Plan Fluor, NA 0.13, 10× (Nikon Plan Apo λ, NA 0.45, FoV of 819 μm), 20× (Nikon CFI Plan Apochromat λ, NA 0.75 and 100× (Nikon Plan Apo VC, Oil Immersion, NA 1.4 were used with an EMCCD camera (iXon Life 897, Andor Technology, Belfast, Northern Ireland) or a sCMOS Hamamatsu Orca Flash 4.0 (Hamamatsu, Shizuoka Pref., Japan). The videos were recorded with frame rates ranging from 10 to 116 $s^{-1}$. For the polarization measurement, the emitted fluorescence light was passed through an emission light splitter (OptoSplit II, Cairn Research Ltd., Faversham, UK). Two perpendicular linear polarization channels were set up at a ± 45 ° angle to the long axis of the array. The ambient temperature in the close proximity of the device was measured to 22 ± 1.4 °C (where ± denotes standard deviation). A schematic of the device and photo of a typical experimental setup can be seen in Figure 9.

*Flow rate measurements and flow velocity calculation*

The volumetric flow rate of the liquid, $Q$, was controlled by applying a pressure difference across the channel, $\Delta p$. The volumetric flow rate of the liquid, $Q$, follows from the Hagen-Poiseuille law, $\Delta P = R_h Q$, where $R_h$ is the hydraulic resistance.

The mean flow velocity inside the array, $u$, which is given in the figures is estimated based on $u \approx Q/A$, where $A$ is the cross-sectional area at an array row, $A = h \times w \times N$, where $h$ is the depth of the device, $w$ is the gap between the pillars and $N$ is the number of gaps per row and $Q$ is the measured volumetric flow rate.

We would like to note that the cross section of the deformable PDMS device expands as a function of the pressure difference between the interior and exterior, which leads to a small decrease in $R_h$, primarily at the high-pressure side of the device. This is seen via a non-linear dependence of $Q$ on $\Delta p$. Figure 10 shows this relationship with varying DNA concentration, solvent viscosity, ionic strength, and polymer length. It is also reasonable to expect the onset of any viscoelastic phenomena in the device to affect energy dissipation and that this will also affect the flow rate as a function of the applied pressure [59]. These effects are not taken into account in the present work and are considered to be small.

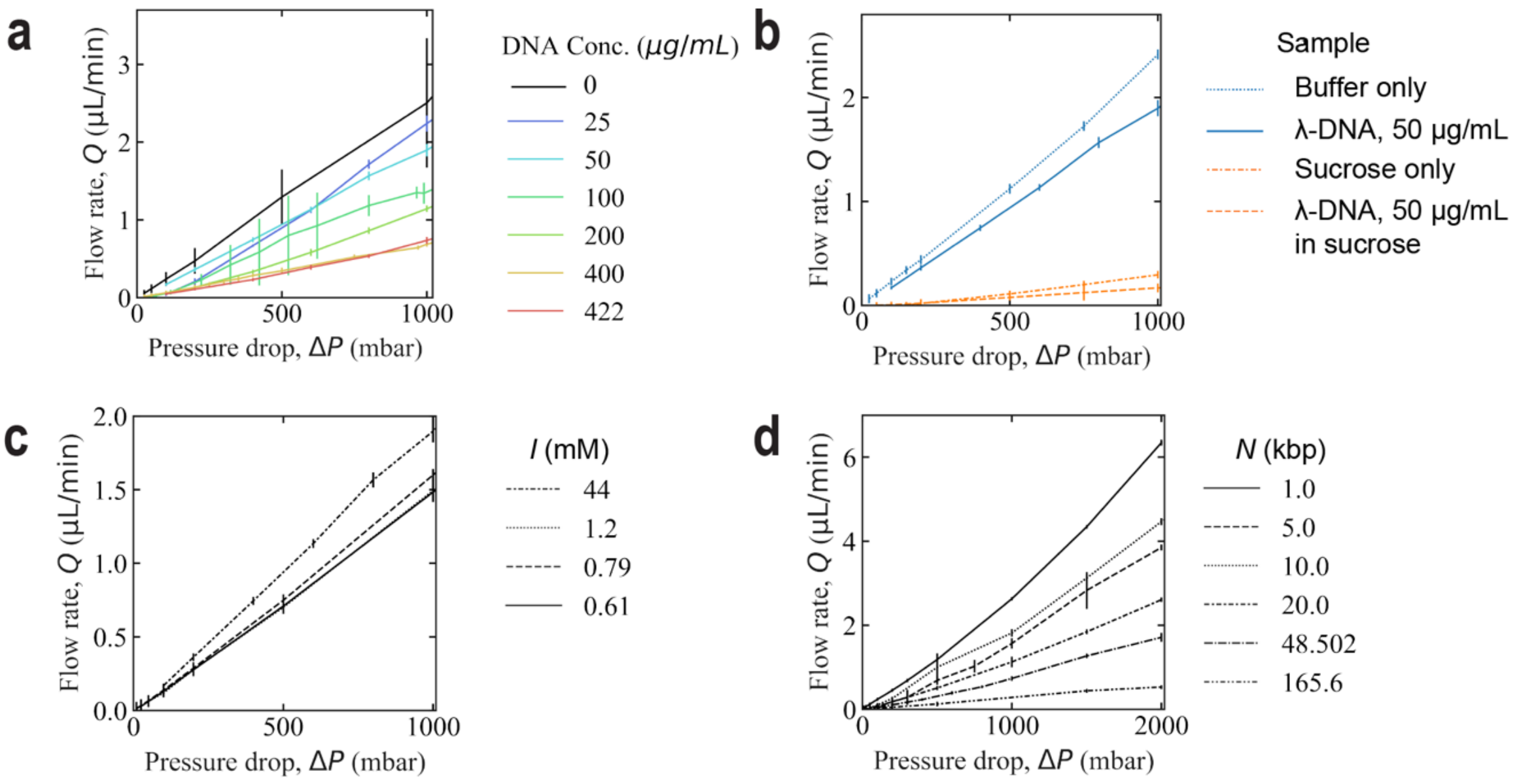

**Figure 10.** Volumetric flow rate dependence on the applied pressure drop across the device for varying DNA concentrations (a), fluid composition (b), ionic strength (c), and polymer length (d). In (a and c), λ DNA is used for all samples. In (d), the same concentration of 422 µg/mL is used for all lengths. 5× TE and 3% BME ($I$ = 44 mM) is used as a buffer in (a, b and d). The buffers used in (c) can be read from Table 1. The error bars denote the standard deviation. The number of mean flow rate data points are 103, 32, 37, and 35, for each plot, respectively. The flow rate and pressure for each data point has been recorded during a duration of at least 20 s.

### *Sample preparation*

All DNA solutions, 1 kbp, 5 kbp, 10 kbp and 20 kbp (NoLimits, 500 µg/mL, Thermo Fischer Scientific, Waltham, MA, USA); lambda phage DNA (λ DNA, 48.502 kbp, New England Bio-labs, Ipswich, MA, USA); and T4 DNA (T4GT7, 165.6 kbp, Nippon Gene, Tokyo, Japan) diluted to various concentrations (between 50 µg/mL and 422 µg/mL) in filtered (200 nm pore) 5× Tris EDTA (50 mM Tris-Hcl and 5 mM EDTA, pH 8) and stained 200:1 with the bisintercalating dye YOYO-1 iodide (491ex/509em, Life Technologies, Carlsbad, CA, USA) for 2 h at 50 °C. Low-salt (0.1× TE, 0.13× TE, and 0.2× TE) samples were in 1× Tris EDTA. Both low salt samples and all samples measured at low magnification (2× and 4×) or higher magnification (20× and 100×) were stained at 10:1 or 50:1 to increase the signal to noise ratio. 3% (v/v) β-mercaptoethanol (BME) was added to minimize photo-bleaching and 0.001% (w/w) Pluronic® F-127 (10 µg/mL and ~12,500 Da per polymer) to prevent non-specific sticking of DNA or dye molecules to the channel walls (except when noted otherwise). For λ DNA samples, the solution was heated to 65 °C for 10 min and then rapidly cooled in an ice bath to remove concatemers. **Caution!** β-mercaptoethanol is fatal in contact with skin, toxic if inhaled or swallowed, may cause organ damage, causes serious eye damage, and is very toxic to aquatic life with long lasting effects. Avoid breathing in any form of the chemical. YOYO-1 causes mild skin irritation, eye irritation and may cause respiratory irritation. Both these substances must be handled with care and appropriate personal protective equipment.

**Table 1.** Physical properties of the λ DNA samples.

| Figure No.[a] | $C$ (μg/mL) | $L$ (μm) | Dye:bp[b] | Buffer | $I$ (mM) | $l_p$ (nm) | $w_{eff}$ (nm) | $R_g$ (μm) | $\tau_Z$ (s) | $C/C^*$ |
|---|---|---|---|---|---|---|---|---|---|---|
| 6a, 6b | 422 | 17 | 1:200 | 5× TE, 3% BME | 44 | 50.7 | 4.63 | 0.47 | 2.1 | 3.6 |
| 4, 5, 6a | 400 | 17 | 1:200 | 5× TE, 3% BME | 44 | 50.7 | 4.63 | 0.47 | 2.1 | 3.4 |
| 2, 3 | 400 | 19 | 1:10 | 5× TE | 31 | 51.0 | 5.79 | 0.54 | 2.6 | 5.1 |
| 1 | 400 | 17 | 1:50 | 1.8× TE | 11 | 52.9 | 11.2 | 0.62 | 2.7 | 7.4 |
| 6a | 200 | 17 | 1:200 | 5× TE, 3% BME | 44 | 50.7 | 4.63 | 0.47 | 2.1 | 1.7 |
| 6a | 100 | 17 | 1:200 | 5× TE, 3% BME | 44 | 50.7 | 4.63 | 0.47 | 2.1 | 0.86 |
| 6a, 6c | 50 | 17 | 1:200 | 5× TE, 3% BME | 44 | 50.7 | 4.63 | 0.47 | 2.1 | 0.43 |
| 6a | 25 | 17 | 1:200 | 5× TE, 3% BME | 44 | 50.7 | 4.63 | 0.47 | 2.1 | 0.21 |
| 6c | 50 | 17 | 1:200 | 0.20× TE | 1.2 | 76.6 | 43.3 | 0.88 | 4.2 | 2.7 |
| 6c | 50 | 17 | 1:200 | 0.13× TE | 0.79 | 91.0 | 56 | 0.80 | 3.4 | 2.1 |
| 6c | 50 | 17 | 1:200 | 0.10× TE | 0.61 | 103 | 65.4 | 0.93 | 5.0 | 3.2 |

For all entries we have $T$ = 22 °C. [a] The column "Figure No." refers to the corresponding figure the sample appears in. [b] Dye:base pair staining ratio.

**Table 2.** List of physical properties for the polymer length samples at high salt.

| Figure No.[a] | Sample | N (kbp) | $L$ (μm) | $R_g$ (μm) | $\tau_Z$ *(s)* | $C$* (μg/mL) | C/C* |
|---|---|---|---|---|---|---|---|
| 6b | 1 kbp | 1 | 0.30 | 0.05 | $2.2 \times 10^{-3}$ | 2800 | 0.15 |
| 6b | 5 kbp | 5 | 1.7 | 0.13 | $37 \times 10^{-3}$ | 670 | 0.65 |
| 6b | 10 kbp | 10 | 3.4 | 0.19 | 0.13 | 390 | 1.1 |
| 6b | 20 kbp | 20 | 6.9 | 0.28 | 0.43 | 230 | 1.9 |
| 6b | λ DNA | 48.5 | 17 | 0.47 | 2.1 | 120 | 3.6 |
| 6b | T4 DNA | 166 | 57 | 0.98 | 18 | 46.0 | 9.2 |

Values based on a buffer of 5× TE and 3% BME at $T$ = 22 °C and with a dye to base pair ratio of 1:200 for the DNA. $C$ = 422 μg/mL. $I$ = 44 mM and thus $l_p$ = 50.7 nm and $w_{eff}$ = 4.63 nm. [a]The column "Figure No." refers to the corresponding figure the sample appears in.

Note that the relative concentration of Pluronic® F-127 is much lower for the concentrated DNA samples (approx. 40× for 400 μg/mL DNA). In addition, it is much less than $C$* for Pluronic® (the molecular weight of Pluronic® F-127 is approximately 2500× less than for λ DNA). Moreover, the molecules are believed to be bound to the surfaces to a large extent further decreasing the bulk concentration of the Pluronic®. See Supporting Tables S1 and S2 for a detailed comparison between λ DNA and Pluronic® polymers.

The sucrose samples were prepared by first adding powdered sucrose (molecular biology grade, Sigma-Aldrich, Saint Louis, MO, USA) to water and dissolving it under vigorous stirring at 50 °C. The sucrose solution was added to stained DNA solutions to reach a sucrose concentration of 43.4% (w/w).

*Image processing of the pillar-free images*

This section describes the image processing for the images and kymographs in the main section in which the pillars have been removed. This type of processing was only performed on magnifications using the 10× objective [see Figure 11(a and b)]. The regions making up the pillar-free images were defined to cover the area between the pillars in the quadratic array, *i.e.* 3 × 2 pixels (4.8 μm × 3.2 μm), see Figure 11(c). The mean intensity of these regions corresponds to each pixel value in the pillar-free images. The region center coordinates were automatically identified as half-way in-between the centers of the pillars. The pixels defining the pillars were themselves located by applying the Sobel edge detection algorithm followed by local Otsu thresholding.

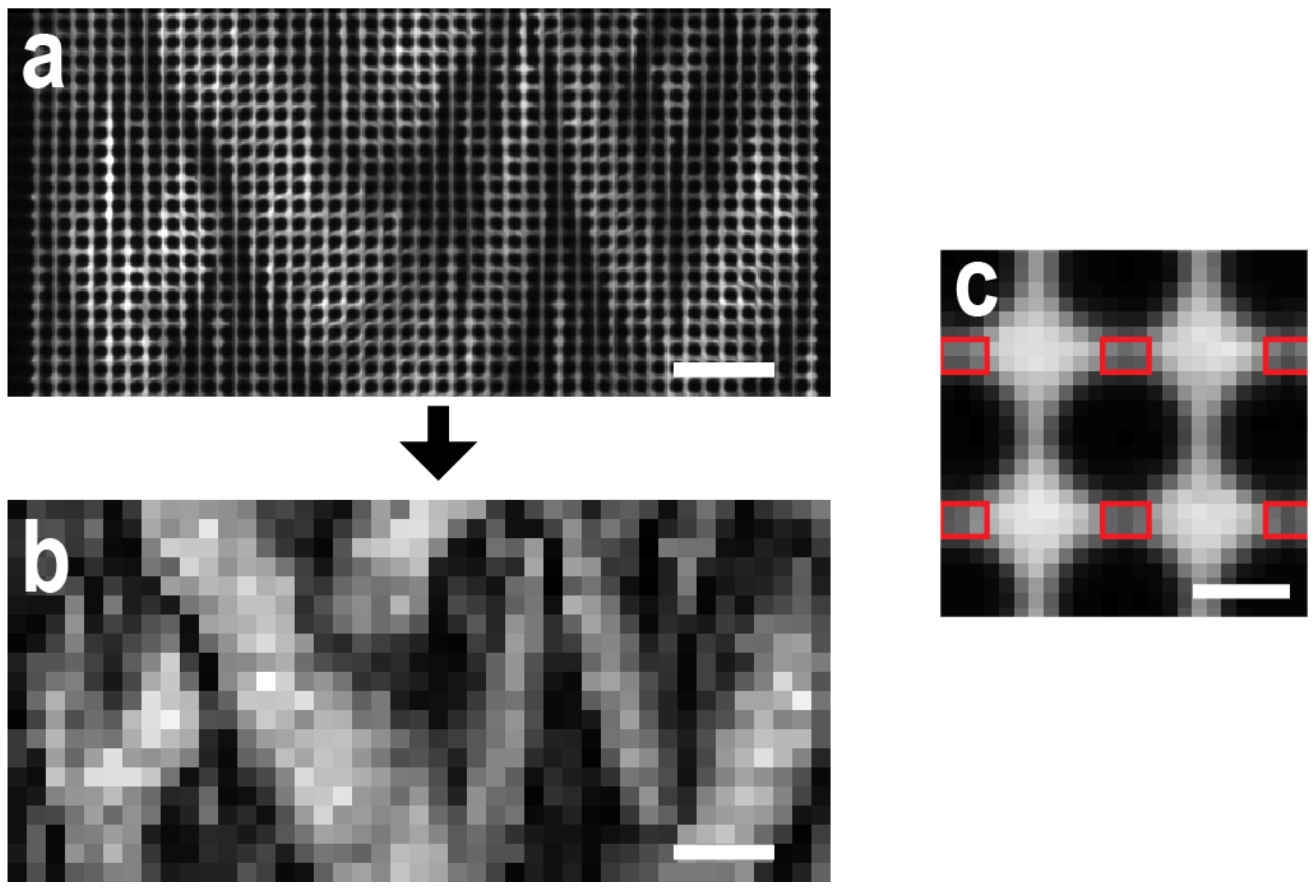

**Figure 11.** Overview of the pillar removal process. A micrograph (a) is converted into a pixel array (b) where the mean value inside each region between the pillars in the flow direction constitutes each pixel value. The regions are denoted by the red rectangles in (c). Scale bars are 100 μm for (a) and (b), and 10 μm for (c).

*Two-channel polarization image processing*

The emission light was split into two perpendicular polarizations and projected onto separate halves of the camera sensor using an Optosplit II. These halves were then overlaid using an Enhanced Correlation Coefficient (ECC) image alignment algorithm from the OpenCV package in Python. The hue, saturation and value (HSV) color model was then employed to visualize the total fluorescence signal together with the polarization signal in the same image, see Figure 12. The total fluorescence intensity of the two polarization halves is represented by the pixel value whereas the polarization emission ratio, *P,* represents the hue in the model. The limits of *P* are set to -0.2 and 0.2, translated to 0 (red) and 1/3 (green) in the HSV model. The saturation level is set to 1 for all pixels, not containing any information about the fluorescent micrographs.

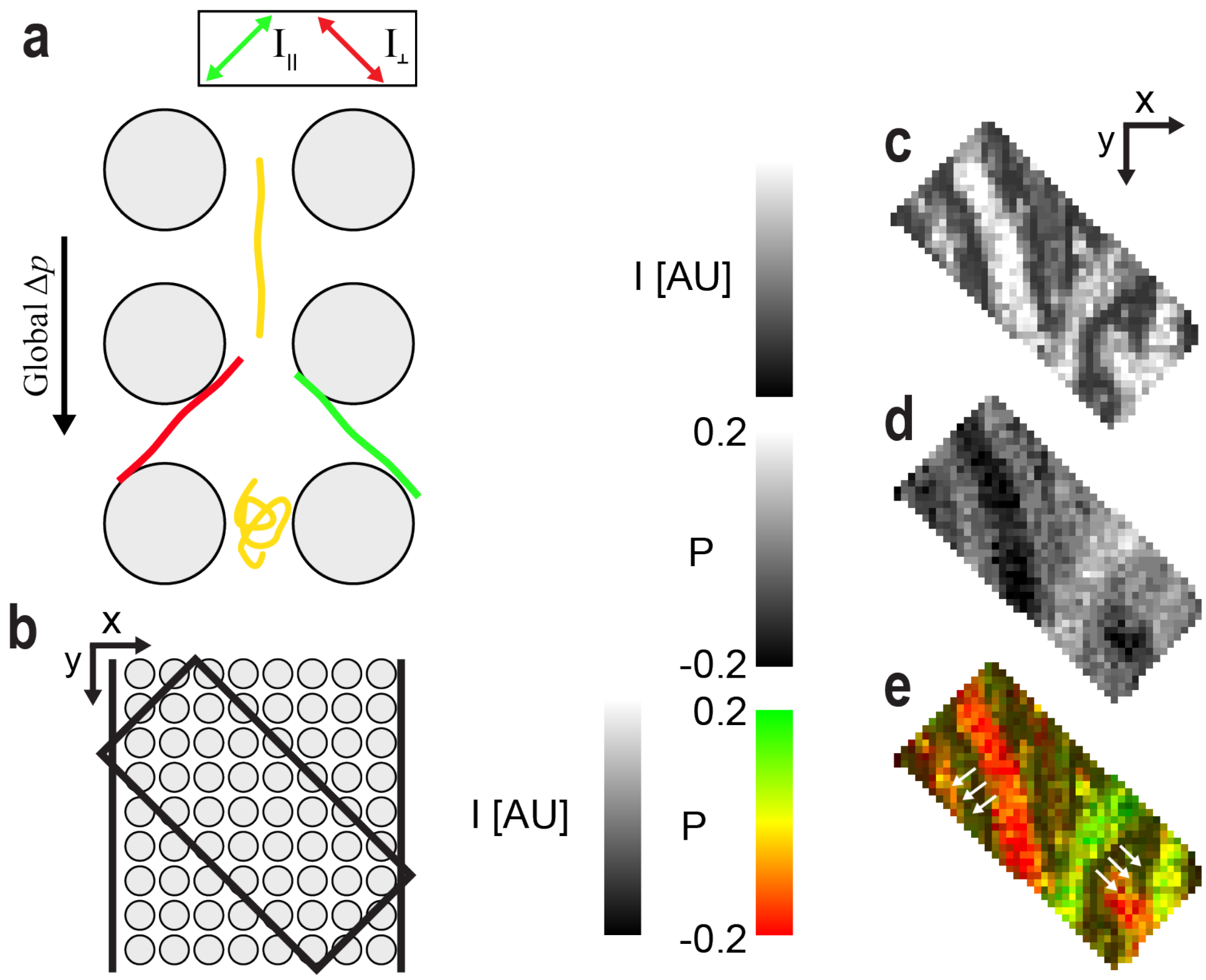

**Figure 12.** Illustration of the polarization image processing. (a) Schematic showing the colors used for different conformations of the DNA molecules. Note that polarization microscopy with a fixed orientation of the polarizer cannot distinguish between the case of a randomly coiled DNA and a DNA that is stretched in a direction that is right in between the two perpendicular polarization directions that are detected. (b) The tilted rectangle is a schematic of the camera field of view over the array for the processed micrographs in (c-e). (c) displays the fluorescence intensity, $I$, (d) the polarization emission ratio, $P$, and (e) a combination of both, where I is represented by the pixel value and P is represented by the hue according to the hue, saturation and value (HSV) color model.

## ASSOCIATED CONTENT

### Supporting Information

Additional relevant details as follows (PDF): wave results of adding sucrose to a low concentration sample, calculation of the overlap concentration and the ionic strength, calculations of various dimensionless numbers, comparison of polymer weight and concentration between Pluronic® and λ DNA, rheology measurements, detailed description of the supporting movies (Movies S1-S5).

### Author contributions.

The author contribution statements have been written according to CRediT (Contributor Roles Taxonomy, see credit.niso.org for role descriptions): Conceptualization, O.E.S., J.P.B., and J.O.T.; Data Curation, O.E.S; Formal Analysis, O.E.S.; Funding Acquisition, J.O.T.; Investigation, O.E.S., and J.P.B.; Methodology, O.E.S., J.P.B., and J.O.T.; Project Administration, J.O.T.; Resources, J.O.T.; Software, O.E.S.; Supervision, J.P.B., and J.O.T.; Validation, O.E.S., J.P.B., and J.O.T.; Visualization, O.E.S., J.P.B., and J.O.T.; Writing - Original Draft, O.E.S., J.P.B., and J.O.T.; Writing - Reviewing & Editing, O.E.S., J.P.B., and J.O.T.; O.E.S discovered the waves.

### Conflicts of Interest

There are no conflicts to declare.

### Acknowledgements

This research was funded by the European Union within Horizon2020, grant number 634890 (project name BeyondSeq), EuroNanoMed (project name NanoDiaBac), by the Swedish Research council, grant number 2016-05739 and NanoLund. All device processing was conducted within Lund Nano Lab.

## Data Availability

The data presented in this study are openly available in Harvard Dataverse:

Ström, O. E., Beech, J. P. & Tegenfeldt, J. O. Short and Long-range cyclic patterns in flows of DNA solutions in microfluidic obstacle arrays. Harvard Dataverse. (2022) https://doi.org/10.7910/DVN/OMCXZG.

**Supporting Information for**

# Short and Long-range cyclic patterns in flows of DNA solutions in microfluidic obstacle arrays

Oskar E. Ström, Jason P. Beech, Jonas O. Tegenfeldt(*)

Division of Solid State Physics, Department of Physics, Lund University

NanoLund, Lund University.

*Jonas O. Tegenfeldt.

**Email:** jonas.tegenfeldt@ftf.lth.se

This PDF file includes:

Supporting text

Figures S1 to S5

SI References

Other supporting information for this manuscript include the following:

Movies S1 to S7

## 1. Addition of sucrose to the DNA sample - Boger fluid

To explore the effect of shear thinning, we prepare a Boger fluid, a fluid with elastic properties but without shear thinning. Figure S1 shows clearly that the added sucrose results in wave formation with a DNA concentration that would otherwise be insufficient.

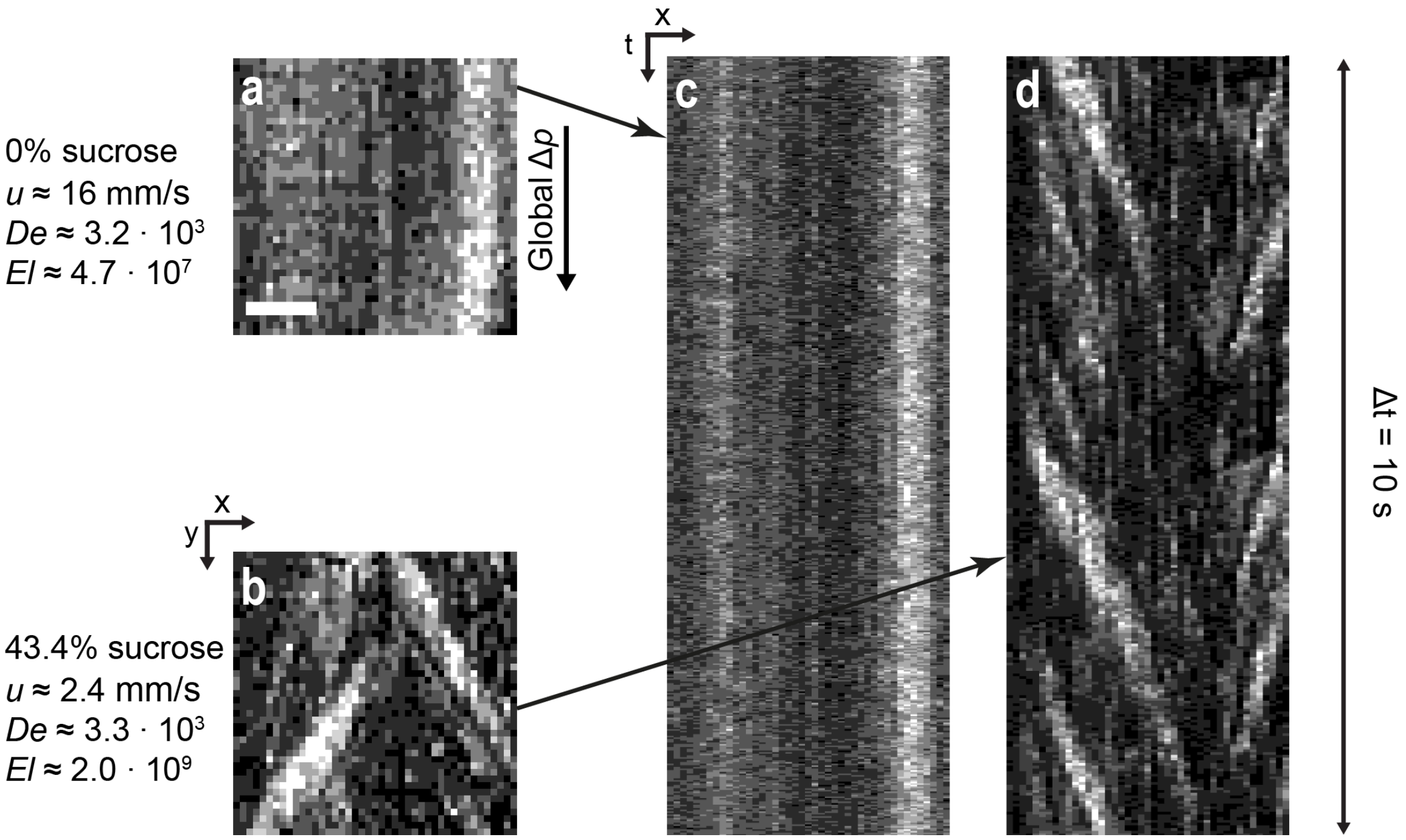

**Fig. S1.** Low DNA concentrations (50 µg/mL, C/C* = 0.43, λ DNA) without (a, c) and with (b, d) added sucrose (w/w %) in the quadratic array. (a) and (b) show snapshots or arrays (pillars removed by image processing, see materials and methods). (c) and (d) show the corresponding kymographs at the middlemost row (y direction) with the same horizontal spatial scale (x direction). The zero-shear viscosity of the sample with sucrose is approximately four times higher than that without (6.5 mPas and 1.9 mPas respectively, based on data from [1] and [2]). The infinite-shear viscosity is however about 5.4× higher as the viscosity of DNA solutions have been shown to approach the solvent viscosity at very high shear rates [1]. The spatial scales are identical for all images with the scale bar representing 200 µm.

## 2. Calculation of the overlap concentration and the ionic strength

The overlap concentration [3] is given by $C^* = M/[(4\pi/3)R_g^3 N_A]$. Here, $M$ is the molecular weight of the DNA and $N_A$ is Avogadro's number. $C^*$ is the concentration above which DNA molecules no longer behave like isolated, individual molecules. Together with the expression for the radius of the polymer below, we obtain an expression for $C^*$ that depends on the contour length, $L$, as $C^* \propto L^{(1-3\nu)} \approx L^{-0.8}$, where $\nu$ is the Flory exponent, $\nu = 0.5877$ [4]. Fig. S2 shows the relationship of the overlap concentration with $I$ and $L$. We illustrate the dependence of the overlap concentration on ionic strength and on molecular length in Fig. S2.

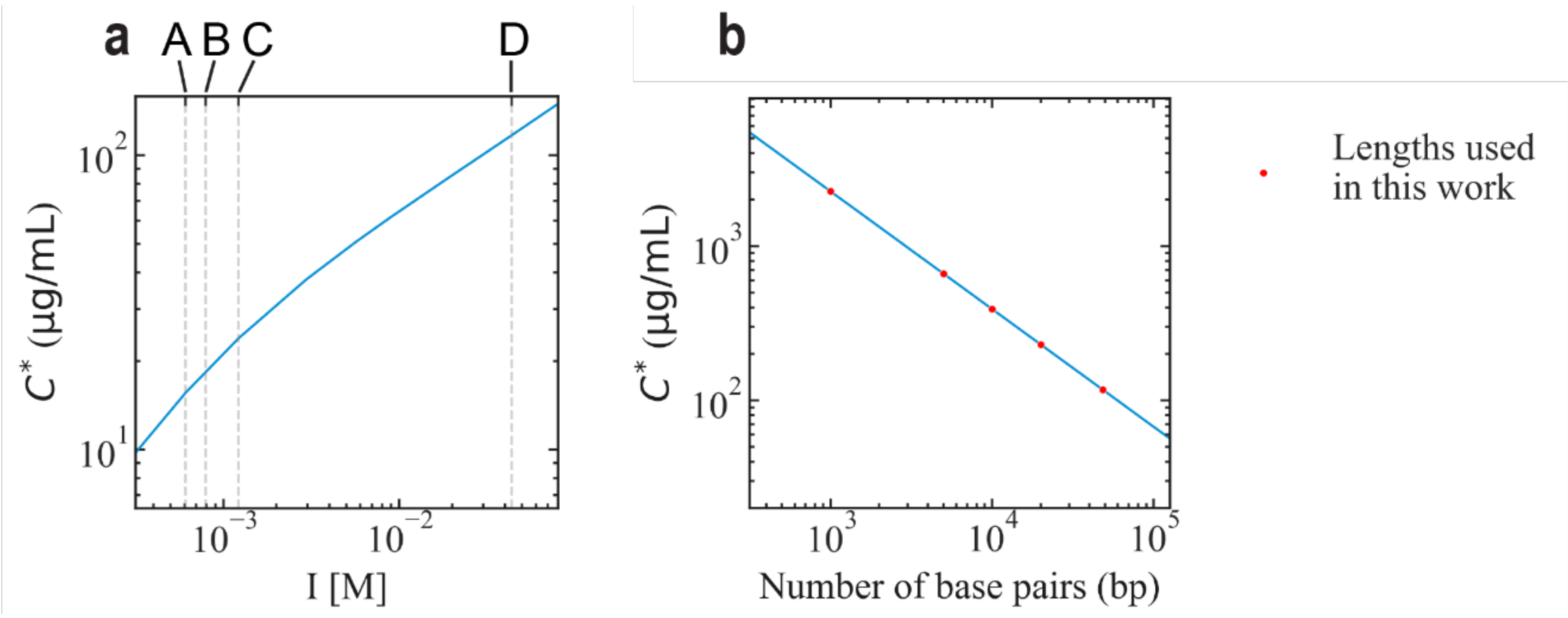

**Fig. S2.** The dependence of the overlap concentration, $C^*$, on the ionic strength for λ DNA (a) and the number of base pairs of the DNA (b) at high salt (5× TE, 3% BME). Values are based on T = 22 °C and with a dye to base pair ratio of 1:200. The buffer concentrations used for our work are indicated by the dashed vertical lines in (a) as 0.1× TE, 0.13× TE, 0.2× TE, and 5× TE and 3% BME for A-D, respectively. The lengths of the DNA used for our work are highlighted using red dots in (b).

The radius of gyration, $R_g$, used in our calculations for $C^*$ is estimated using the worm-like chain (WLC) model (also known as the Kratky-Porod model) as $R_g \approx R_e/\sqrt{6}$, where $R_e$ is the polymer end-end distance. To make the estimate as accurate as possible, electrostatic interactions and excluded volume effects are taken into account. $R_e$ is then described as [5]:

$$R_e \approx \left(w_{eff} l_p\right)^{1/5} (bN)^{\nu} \qquad [1]$$

where $w_{eff}$ is the effective width of the polymer, $l_p$ is the persistence length, $b = 2l_p$ is the Kuhn length and $N$ is the number of Kuhn segments, $N = L/b$, where $L$ is the contour length. For every dye molecule incorporated into the DNA strand, the contour length is increased by 0.51 nm [6]. $l_p$ depends on ionic strength according to Odijk−Skolnick−Fixman (OSF) theory[7-9] as:

$$l_p = l_p' + \frac{0.0324\text{M}}{I}\,\text{nm} \qquad [2]$$

where $l_p' = 50$ nm is the bare persistence length and $I$ is the ionic strength of the buffer. See below for how we calculated $I$ for the different buffers used.

The effective width for strongly charged chains, such as DNA, is given by [5]:

$$w_{eff} = \frac{1}{\kappa}\left[0.7704 + log\left(\frac{\nu_{eff}^2}{2\epsilon\varepsilon_0 k_b T\kappa}\right)\right] \qquad [3]$$

where $1/\kappa$ is the Debye length, $\epsilon$ the dielectric constant of water, $\varepsilon_0$ the permittivity of free space, $\nu_{eff}$ is the effective DNA line charge density, $k_B$ is Boltzman's constant and $T$ the temperature. We use $\nu_{eff} = -0.593$ $e$/Å [10], where $e$ is the elementary charge. For a buffer containing 5× TE and 3% BME ($I$ = 44 mM) at $T$ = 22 °C we calculate $w_{eff} = 4.6$ nm and for a buffer of 1× TE at the same T, $w_{eff} = 43$ nm which corresponds well with literature values from Iarko *et al.* [11].

The Ionic strength, *I*, is calculated by summing the product of the concentration and squared charge of all ions in the solution:

$$I = \frac{1}{2}\sum_{i=0}^{n} z_i^2 c_i \qquad [4]$$

where the one half is added to include both anions and cations and $c_i$ is the concentration of an ionic species with the valence $z_i$. We use the python library SymPy to compute *I* based on the detailed description by Iarko *et al.* [11]. The equilibrium concentrations are calculated by solving a series of equations consisting of the equilibrium conditions based on the law of mass action. The activity coefficients of these conditions are calculated using the Davis equation. Because the activity coefficient depends on the ionic strength which in turn depends on the activity coefficients, the system of equations is iterated until a stable solution is found. The initial activity coefficients are set to 1. The percent error (100 × [calculated value – literature value] / literature value) is below 0.6% [12].

**3. Calculations of various dimensionless numbers**

Dimensionless numbers are used extensively in our description of our work. However, it is important to be aware that they vary across the devices that we use both in space and in time. The main purpose is to give an overall idea of the nominal properties of the fluids and what factors influence the behavior of the fluids in the devices.

The Reynolds number describes the relationship between inertial effects and viscous effects, and is calculated according to $\mathrm{R}e \equiv \rho \cdot u \cdot w/\eta_s$,where *u* is the mean fluid velocity in the pillar gaps of the array, $w$ is the gap width between the pillars, $\rho$ is the fluid density and $\eta_s$ is the solvent viscosity. In our case we have $Re \ll 1$ and thus we can treat any inertial effects as negligible.

The Deborah number, *De,* describes the ratio between the relaxation time of the system (here the DNA molecules) and the time scale of the applied forces (here the interaction time between the flowing molecules and individual pillars) [13]. In the present work, we define $De \equiv (u/L_{pp})\tau_Z$, where, $L_{pp}$ is the center-to-center distance between array rows and $\tau_Z$ is the Zimm relaxation time of the polymer. Note that $\tau_Z$ only gives approximate values of the relaxation time as it assumes the conditions to be ideal, the solution to be dilute and the solution to be in equilibrium.

We define the elasticity number, $El = De/Re$, which describes the ratio of elastic stress to inertial stress, with the Deborah number rather than the Weissenberg number since the shear rate in our system is not constant along the channel.

We estimate the relaxation time, $\tau_z$, using the Zimm relaxation time [14], $\tau_{Zimm} = R_g^2/D_Z \approx \eta_s\, b^3 N^{3\nu}/(k_B T) \propto L^{3\nu} \propto L^{1.76}$ where $\eta_s$ is the solvent viscosity, *b* is the Kuhn length, $k_B$ is Boltzmann's constant, *T* is temperature, *N* is the number of Kuhn segments and $\nu$ is the Flory exponent. In the semidilute regime, the equilibrium relaxation time can be expressed in terms of the ratio between the concentration and the overlap concentration [15]:

$$\tau \propto (C/C^*)^{(2-3\nu)/(3\nu-1))} \approx (C/C^*)^{0.31}.$$

The magnitudes of the dimensionless numbers presented in the main text should be considered lower bounds. Based on data from [16-19] we can estimate that the relaxation time and thus our estimates for *De* and *El* can be increased by at least a factor of 3 for $C \approx 4C^*$ with concentrated λ DNA samples and a factor of 30 for the concentrated T4 DNA sample ($C \approx 9C^*$). The shear thinning has a two-fold decreasing effect on *El* as both the viscosity [1,17] and the relaxation time [20] have been shown to reduce with higher shear rates. We would ultimately expect *El* to be significantly higher than reported in this work at high *C* and low flow velocities and slightly lower at high flow velocities.

**Table S1.** Comparison of polymer weight and concentration between Pluronic® and λ DNA.

| Polymer | Polymer weight (MDa) | *C* (μg/mL) | *C* (% w/w) |
|---|---|---|---|
| Pluronic® F-127 (poloxamer 407) | 0.012 | 10 | 0.001% |
| lambda phage DNA (λ DNA, 48.5 kbp) | 31.5 | 400 | 0.04% |

**Table S2.** Weight and concentration ratios between λ DNA and Pluronic®.

| Weight Ratio(λ DNA/Pluronic®) | Concentration Ratio (λ DNA/Pluronic®) |
|---|---|
| 2522 | 39.98 |

## 4. Rheology measurements

The relaxation time of a solution of 400 μg/mL lambda phage DNA in 5× TE buffer was measured using a stress-controlled rheometer (Physica MCR 301, Anton Paar) with a 25 mm cone plate geometry (CP 25-1) at 25 °C. An amplitude measurement was performed to find the linear viscoelastic (LVE) region, Fig. S3, and 30% of the maximum strain was chosen for subsequent frequency measurements. Three measurement series were taken and the means of G' and G'' used to find the cross-over frequency, at which the relaxation time τ is equal to the inverse of the angular frequency. This gave a value for the relaxation time of 1.43 s, Fig. S4, S5. This relaxation time is within the same order of magnitude as the calculated Zimm relaxation time for the same solution (2.6 s, see Table 1).

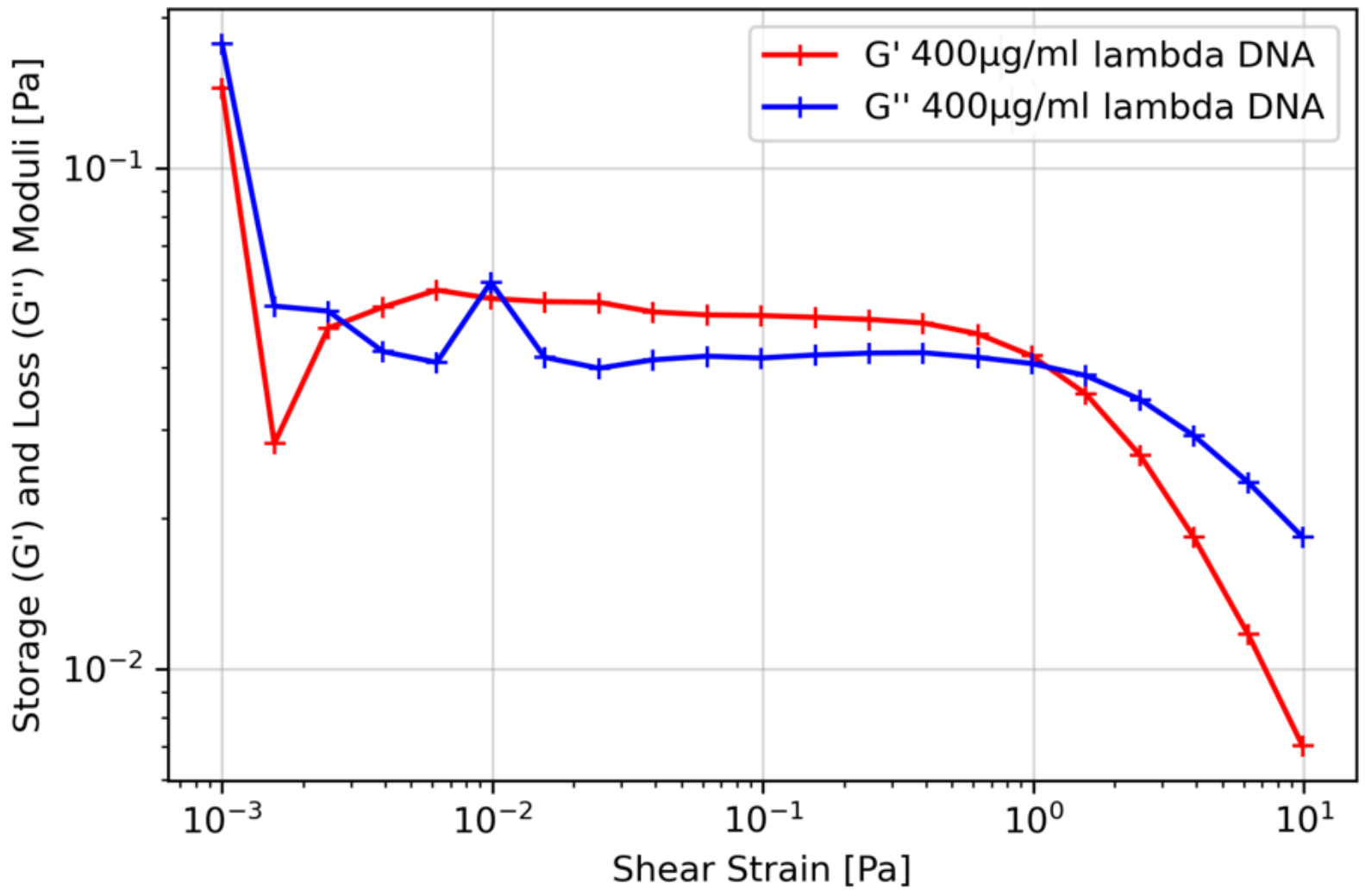

**Fig. S3.** Amplitude sweep to determine the linear viscoelastic (LVE) region.

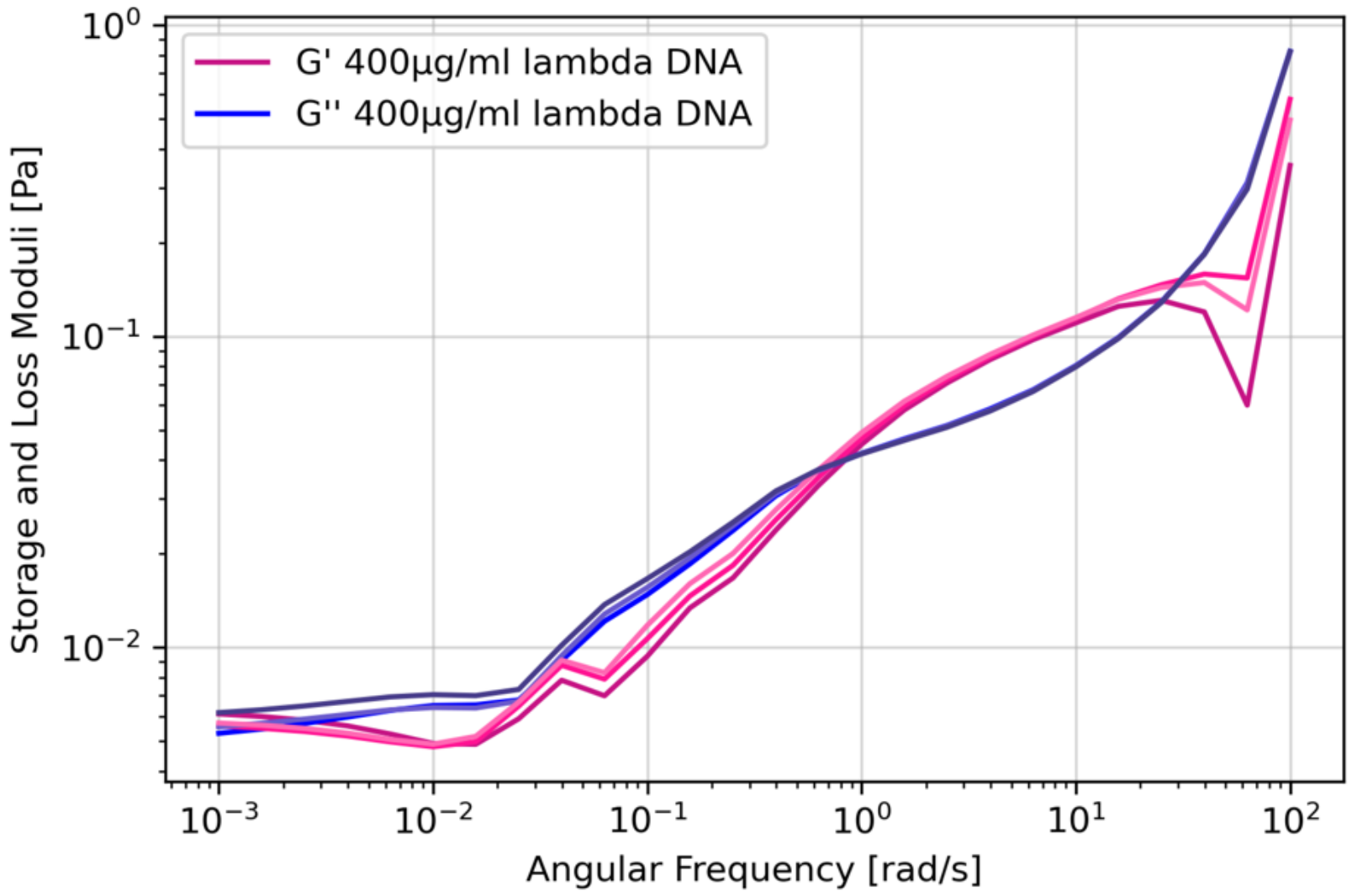

**Fig. S4.** Frequency sweeps (triplicate) for a solution of 400 µg/mL λ DNA in 5× TE buffer.

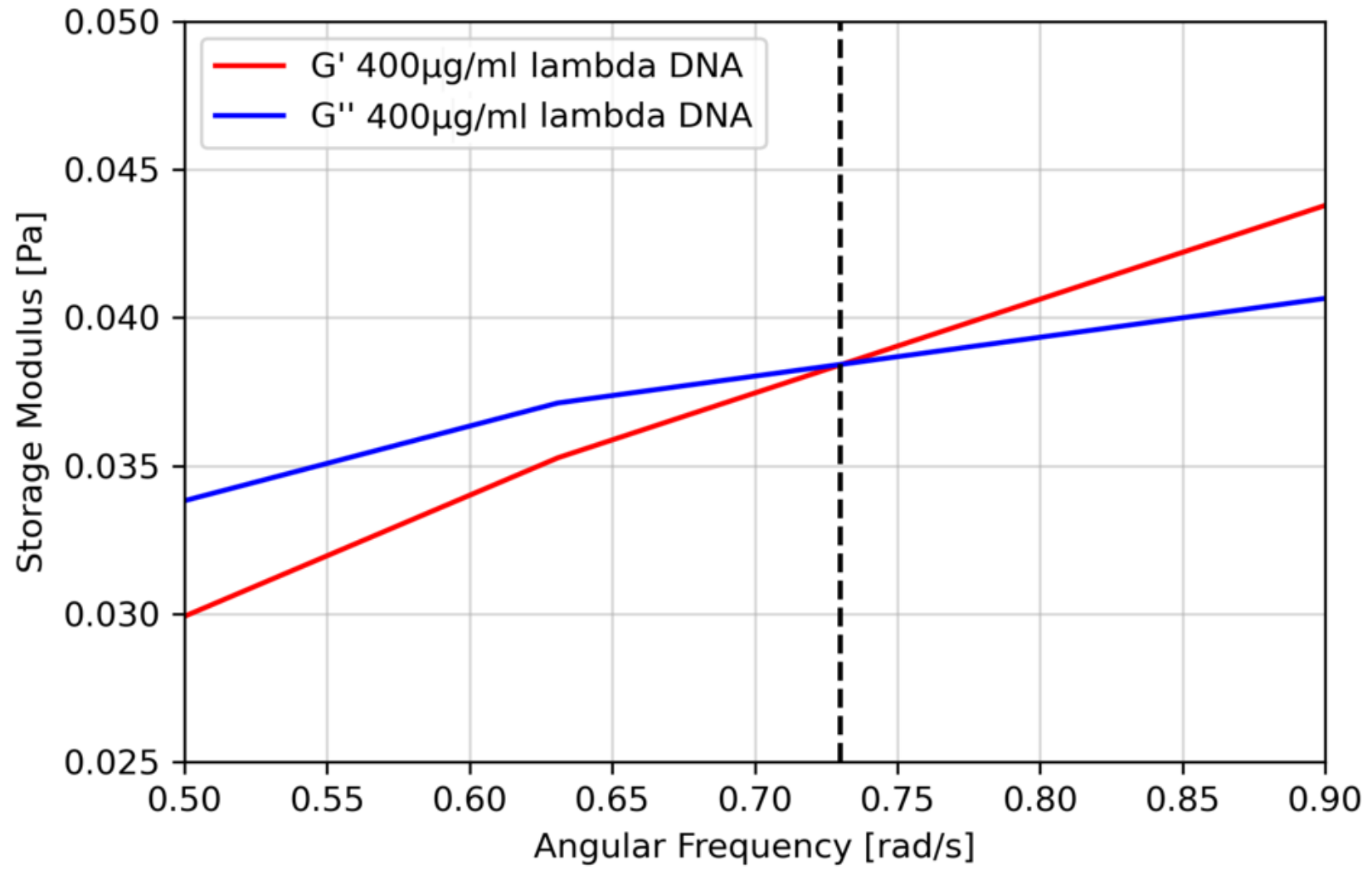

**Fig. S5.** The average of the three frequency sweeps from Fig. S4. The relaxation time is derived from the cross-over frequency where G' = G'' which is shown with the black, vertical dashed line.

## 5. Movies

**Movie S1. (separate file).** Low-magnification (2×) videographs comparing low flow velocity ($u \approx 10^2$ µm/s, $De \approx 10^2$) to high flow velocity ($u \approx 10^3$ µm/s, $De \approx 10^3$) of λ DNA solutions flowing through quadratic and disordered arrays. The Movie contains data that corresponds to Fig. 1.

**Movie S2. (separate file)**. Low-magnification (4×) fluorescence videographs where $\Delta p$ is ramped from no flow to high flow rate ($De \approx 10^3$) with a λ DNA solution of $C$ = 400 µg/mL and high salt ($I$ = 44 mM).

**Movie S3. (separate file).** High-magnification (20×) videographs of low the three flow regimes S, C and W with a λ DNA solution of high $C$ = 400 µg/mL and high salt ($I$ = 31 mM). Note that the video playback rate is set the same for all flow velocities, so that the higher flow velocity video sections are slowed down. The Movie contains data that corresponds to Fig. 2 and 3.

**Movie S4a. (separate file).** Low-magnification (10×) fluorescence videographs of low $C$ = 50 µg/mL and high $C$ = 400 µg/mL λ DNA solution, at high flow rate ($De \approx 10^3$) and high salt ($I$ = 44 mM).

**Movie S4b. (separate file).** Low-magnification (10×) fluorescence videographs (where the pillars have been removed) of low $C$ = 50 µg/mL and high $C$ = 400 µg/mL λ DNA solution, at high flow rate ($De \approx 10^3$) and high salt ($I$ = 44 mM).

**Movie S5. (separate file).** Low-magnification (4×) fluorescence videograph sweeping the field of view across the entire array for a λ DNA solution of low salt ($I$ = 1.2 mM) and $C$ = 50 µg/mL, high flow rate ($De \approx 10^3$).

**Movie S6. (separate file).** High-magnification (100×) videographs of low ($De \approx 26$) and high flow rates ($De \approx 1.1 \times 10^3$) for a λ DNA solution of high $C$ = 400 µg/mL and high salt ($I$ = 44 mM). Note that the color represents the polarization emission ratio and the pixel value the total fluorescence intensity. The Movie contains data that corresponds to Fig. 4.

**Movie S7. (separate file).** Low-magnification (10×) videographs of low ($De \approx 26$) and high ($De \approx 1.1 \times 10^3$) flow rates for a λ DNA solution of high $C$ = 400 µg/mL and high salt ($I$ = 44 mM). The pillars have been removed with image processing and each pixel represents the average value of a dead zone. Note that the color represents the polarization emission ratio and the pixel value the total fluorescence intensity. The Movie contains data that corresponds to Fig. 4.